\documentclass[10pt,a4paper,preprint]{elsarticle}
\usepackage{amsmath,amssymb}
\usepackage{graphicx}
\usepackage[utf8]{inputenc}
\biboptions{sort&compress}
\begin{document}
\begin{frontmatter}
\title{Overview of Seniority Isomers}
\author{Bhoomika Maheshwari\corref{mycorrespondingauthor}}
\cortext[mycorrespondingauthor]{Corresponding author}
\ead{mbhoomika.phy@pmf.hr}
\author{Kosuke Nomura}
\address{Department of Physics, Faculty of Science, University of Zagreb, HR-10000, Croatia.}
\begin{abstract}
Nuclear isomers are the metastable excited states of nuclei. The isomers can be categorized into a few classes including spin, seniority, \emph{K}, shape and fission isomers depending upon the hindrance mechanisms. In this paper, we aim to present an overview of seniority isomers, which is a category related to the seniority quantum number. The discussion is mainly based on the concepts of seniority and generalized seniority. Various aspects of seniority isomers and their whereabouts have been covered along with the situations where seniority mixing prevents the isomerism.
\end{abstract}
\begin{keyword}
\texttt{seniority; generalized seniority; isomers}
\end{keyword}
\end{frontmatter}
\section{Introduction}

The nucleus presents a complex quantum system which is still far from being completely understood. Nuclear isomers have been serving as one of the key tools to decipher the nucleonic properties and configurations for more than 100 years~\cite{jain2021,atlas22,walker2022}. The isomers are longer-lived excited quantum states of nuclei. Isomeric applications, in explaining the fundamental structure to adding new dimensions in energy and medical industries, are now very well known. Due to modern-day experimental facilities, a huge amount of experimental data on isomers and their various spectroscopic properties is becoming available each year; for more details, see \textit{The Atlas of Nuclear Isomers}~\cite{atlas22}. The hindrance to the isomeric decays is mostly due to three reasons: different structural matrix elements of initial and final states, small decay energy, and the large change in angular momentum. In simple words, the excited states adopting different configurations from the lower-lying states result in meta-stable states, that is, isomeric states. On the basis of hindrance mechanisms to their respective decays, the isomers can be classified into spin and seniority isomers in spherical regions and \emph{K}, shape and fission isomers in deformed regions, where \emph{K} is the total angular momentum projection on the axis of symmetry for axially deformed nuclei. In this paper, we aim to present the detailed features of seniority isomers and their whereabouts. 

Seniority isomers arise as a result of Racah's seniority selection rules~\cite{racah1943}, which he proposed in 1943 \cite{racah1943} to identify states with the same orbital, spin, and total angular momentum quantum numbers. A quantity introduced only to distinguish states with the same quantum numbers has shown to be so crucial in understanding an entire class of nuclear excitations. From the `vertek', the Jewish word for seniority, \emph{v} is used to denote the seniority quantum number which refers to number of unpaired nucleons, which are not pair coupled to the angular momentum \emph{J} = 0. The concept of seniority was first introduced in the atomic physics. However, the long-range repulsive Coulomb force between electrons pushes low seniority states very high in energy. The seniority was later used in the context of nuclear physics by Flowers~\cite{flowers1952} and Racah~\cite{racah1952} almost simultaneously, where the short-range attractive nuclear force results in a low-lying spectra with lower seniority. This makes the seniority more crucial for the nuclear structure which simplifies the variation of structural properties of the complex nucleonic system, particularly in semi-magic nuclei~\cite{talmi1993,rowe,kota,casten,lawson,heyde,isacker2008,parikh1978}. The seniority quantum number is due to the symmetry in pairing Hamiltonian, and the eigen states of pairing represent the good seniority states. 

\begin{figure}[!ht]
  \includegraphics[width=0.5\textwidth]{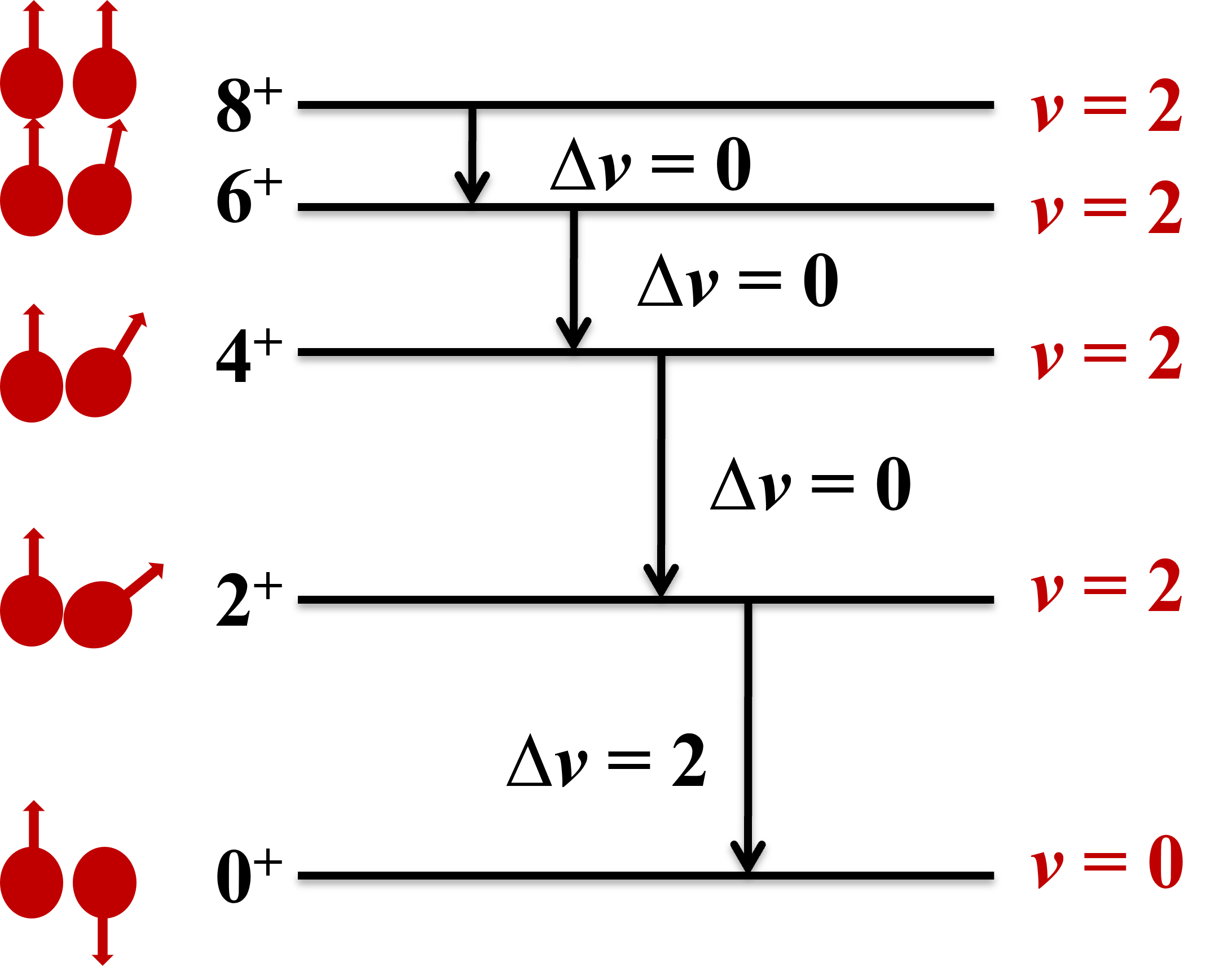}
  \caption{ \label{fig:singlej}(Color online) Schematic level scheme for (9/2)\textsuperscript{2} configuration in single-j shell. The levels are identified by the seniority quantum number on the extreme right. $\Delta v=0$ and $\Delta v=2$ refer to the seniority-preserving and seniority-changing transitions, respectively.}
 
\end{figure}

The seniority in a single-j shell of identical nucleons follow the dynamical symmetry classification of $U(2j+1) \supset Sp(2j+1) \supset  SU(2)$. The states in $j^n$ configuration, \emph{n} particles in the j-shell, constitute a $[1^n]$ irreducible representation of $U(2j+1)$ which further reduces to the $[1^v]$ irreducible representation of $Sp(2j+1)$, with $v=n, n-2, ..., 1,$ or, 0. In other words, $v=1$ is the lowest value for a state with an odd-valence particle, whereas \emph{v} = 0 is the lowest value for even--even nuclei having the ground states as fully paired \emph{J} = 0, \emph{v} = 0. The \linebreak \emph{J} = 0, 2, 4, 6, 8 are the allowed values for \emph{n} = 2 particles in \emph{j} = 9/2. Among which, the states having non-zero total angular momentum, \emph{J} = 2 to 8, have the seniority of \emph{v} = 2 involving one pair-breakup, while \emph{J} = 0 is with seniority \emph{v} = 0. The schematic level scheme for two particles in $j=9/2$ is shown in Figure \ref{fig:singlej}. The lower seniority states generally prefer to lie lower in energy. It is obvious that the seniority mixing is not permitted for two particles in a single-\emph{j} shell with $j\leq{7/2}$, and the \emph{v} is exactly conserved, because there are no two states with the same \emph{J} and seniority \emph{v}. The situation becomes more complex for higher-\emph{j} values where seniority mixing is now possible~\cite{parikh1978,escuderos2006,isacker2008,qi2011}. Seniority may also continue to be a reasonable quantum number for higher-j orbitals, such as \emph{j} = 11/2, 13/2, and so on, but only for a limited set of states.

The good seniority states have some key spectroscopic features such as a constant pairing gap, \emph{n}-independent energies, and the \emph{B} (\emph{E}2) parabolas \cite{isacker2011,maheshwari2016}. In semi-magic nuclei, where only identical nucleons are active, the concept of seniority stays valid and seniority isomerism is likely to arise at the half-filled valence shell due to almost vanishing \emph{B} (\emph{E}2) values. This is due to the direct dependency of matrix elements of electric multipole tensor operator on the $(\Omega-n)$ factor, where $\Omega$ is the pair degeneracy and \textit{n} is the valence nucleon number, leading to vanishing decay rates and moments at the mid-shell. This extra hindrance to decay rates in terms of seniority selection rules leads to the longer half-life of those excited states, which are called seniority isomers.

The validity of seniority for \textit{j} = 9/2 has first been discussed by Lanford in 1969 \cite{lanford1969} and further been extended by Van Isacker \cite{isacker2011} in the context of seniority isomers. For any reasonable nuclear interaction, it has been proven that the off-diagonal interaction matrix element remains small in comparison to the energy difference between two similar $J$ states with different seniority. In ${(9/2)}^4$ configuration, the off-diagonal matrix element is only a few tens of keV in energy, while the two $J = 2^+$ states with $v=2$ and $v=4$ have an energy splitting of about 1 MeV. For the seniority isomers arising from $j=9/2$, it is of more interest to prove the negligible seniority mixing for the $J=4^+$ and $6^+$ states. Among the allowed three states for both $J=4^+$ and $6^+$, two are with $v=2$ and $v=4$ and are situated close in energy to mix easily. One can differentiate the `solvable’ $v = 4$ members of these closely spaced $J = 4^+$ and $6^+$ states as discussed in refs.~\cite{isacker2008,escuderos2006,qi2011}. The only possibility of seniority mixing therefore exists in between the \textit{v} = 2 and the higher-lying $v=4$ states which is again found to be small. For higher-j values, the seniority mixing starts to occur, but one or more of the multiple states still remain of good seniority. This result offers helpful direction for comprehending the seniority isomers resulting from the higher-\textit{j} orbitals, $j>7/2$. Such higher-j orbitals are found in the 50--82 and 82--126 nucleonic valence spaces
and are mostly the intruder and unique-parity orbitals near the shell closures, which are also responsible for isomerism.

There has been a flurry of activities in recent years \cite{sawicka2003,ressler2005,jungclaus2007,lozeva2008,pietri2011,isacker2011,watanbe2013,gottardo2012,astier2012,astier2013,iskra2014,simpson2014,maheshwari2015,maheshwari2016,maheshwari20161,maheshwari2017,kota2017,jain2017,jain20171,maheshwari2019,maheshwari20191,maheshwari2021,maheshwari2022,gorska2022,grawe2021,dobon2021,iskra2016,togashi2018,morales2011}, both experimentally and theoretically, to decipher and search seniority isomers due to its capacities of providing a simpler understanding of the two-body interaction and nucleonic configurations. One of the recent developments is the manifestation of Berry phase in atomic nuclei for the particular case of \textsuperscript{213}Pb using the seniority isomerism \cite{dobon2021}. This paper is not an extensive review for all of such efforts but provides a general overview of the subject in and around semi-magic nuclei on the basis of seniority and generalized seniority. Many microscopic model explanations for these mass regions are also available, but in this overview, we mostly cover the discussion based on the simple concepts of seniority and generalized seniority. The details of quasi-spin algebra for both the single-j and multi-j shells in terms of seniority and generalized seniority, respectively, have briefly been reviewed in Section~\ref{sec2} along with the resulting electromagnetic (plus the seniority) selection rules. Section \ref{sec3} compares the possibility of seniority isomerism in single-j to the generalized seniority isomerism in multi-j. Due to available advanced experimental and theoretical tools, the seniority isomers have been explored at the extremes of the binding energies; a few of them include \cite{jungclaus2007,gottardo2012,watanbe2013,simpson2014,corsi2018}, providing an opportunity to test the rigidity of spherical magic numbers. The examples of seniority isomers along with their structural details are given in Section \ref{sec4}. Section \ref{sec5} reports a collection of predictions where seniority provides a useful guidance. Section \ref{sec6} concludes the paper.

\section{Quasi-Spin Algebra: Single-j and Multi-j}\label{sec2}

Seniority can be expressed simply in terms of quasi-spin operators as defined by  
Kerman~\cite{kerman1961,kerman19611} and Helmers~\cite{helmers1961}. For a single-j shell, a pair creation operator $S_j^+$ that acts on the vacuum state and produces a fully paired-state with \textit{J} = 0 can be written as

\begin{eqnarray}
S_j^+ &=& \sqrt{\frac{2j+1}{2}} A^+(jj;J=0, M=0) \nonumber \\
      &=& \frac{1}{2} \sum_{m} (-1)^{j-m} a_{jm}^{+} a_{j,-m}^{+} \nonumber  \\
	  &=& \sum_{m>0} (-1)^{j-m} a_{jm}^{+} a_{j,-m}^{+} \label{paircreation}
\end{eqnarray} 

Similarly, the pair annihilation operator, a Hermitian conjugate of the pair creation operator, annihilates a pair and can be defined as
\begin{equation}
S_j^- = \sum_{m>0} (-1)^{j-m} a_{j,-m} a_{jm} 
	  \label{pairannihilation}
\end{equation}  
In $j^v$ configuration, the state having maximum seniority $v_{max}$ can therefore be \linebreak defined by 
\begin{equation}
S_j^- |j^v,{v_{max}},J,M \rangle = 0
\end{equation}
There are no more pairs to annihilate. On adding $\frac{n-v}{2}$ pair of particles coupled to \linebreak \textit{J} = 0 to this state, 

\begin{equation}
(S_j^+)^{\frac{n-v}{2}} |j^v,v,J,M \rangle
\end{equation} 

We obtain a state with the same seniority \textit{v}, with $(n-v)$ being the total number of (paired) particles, in the $j^n$ configuration. In the $j^n$ configuration, the states with \emph{n} even or odd have seniorities \textit{v} that are even or odd, respectively. In the configuration $|j^v J>$, there are no particles coupled in pairs to \textit{J} = 0. In addition,

\begin{equation}
[ S_j^{+}, S_j^{-} ] = \sum_{m} {(a_{jm}^+ a_{jm})} - \frac{1}{2} (2j+1)
= {\hat{n}_j} - \Omega = 2 S_j^0.
\end{equation}

Since the SU(2) Lie algebra that these operators follow is similar to the angular momentum (spin) algebra, the scheme with the operators $S_j^+$, $S_j^-$ and $S_j^0$ has been given the name quasi-spin scheme \cite{talmi1993,rowe,casten,lawson,heyde}. In terms of seniority quantum number, Racah \cite{racah1943} introduced a special pairing Hamiltonian, $2 S_j^+ S_j^-$ representing a constant pairing. The good seniority states may be obtained as the eigen states of this pairing Hamiltonian,
\begin{eqnarray}
H_{pair}&=& -2 G S_j^+ S_j^- \nonumber \\
 &=& -(2j+1)G A^+(jj; J=0, M=0) A (jj; J=0, M=0)
\end{eqnarray}
where \textit{G} is the pairing strength. The pairing eigen values can be obtained as
\begin{eqnarray}
H_{pair}(n,v)&=& -G \Bigg( 2s(s+1)-\frac{1}{2} {(\Omega-n)(\Omega+2-n)} \Bigg) \nonumber\\
&=&-G \frac{n-v}{2} {(2\Omega+2-n-v)} \label{p.e.}
\end{eqnarray}
where quasi-spin $s=\frac{1}{2}(\Omega-v)$ and $S_j^0=\frac{1}{2}{(n-\Omega)}$, the number operator $ n={\hat{n}_j}$, and pair degeneracy (total number of possible pairs) $\Omega=\frac{1}{2} (2j+1) $. The eigen values in Equation~(\ref{p.e.}) correspond to the eigen states with seniority \textit{v} in $j^n$ configuration having $\frac{n-v}{2}$ number of pairs. For the limiting value of $\Omega$ as $\frac{n}{2}$, the pairing energy value becomes zero for $v=2$ states. For a fully paired $J=0$, $v=0$ state, the eigen value is -\textit{nG}, i.e., $-(2j+1)G$. \linebreak The middle of the j-shell is represented by $v= \Omega$, which is also the maximum allowed value of $v$. For example, in the $f_{7/2}^4$ configuration, the maximum allowed spin is $8^+$, which can only be obtained with maximally aligning the $m$-components of all four particles in $f_{7/2}$, as allowed by the Pauli principle. The corresponding seniority of such a $J=8$ state is $v=4$, which means that no particles are pair-coupled to $J=0$. As a consequence, the $0^+$ fully pair-coupled states are found as the ground states in even--even nuclei. It should be noted that this description works well for the case of identical nucleons, but it may fall short of isospin $(T=0)$ and proton--neutron residual interactions if both kind of nucleons (i.e., protons and neutrons) become active. The semi-magic nuclei, where either protons or neutrons are valence particles with the other being at a closed shell, can be a good example to test the seniority arguments. The maximum pairing and lowest seniority states exist in these nuclei. This may result in a seniority-like the energy band structure shown in Figure \ref{fig:singlej} (in a complete contrast with the deformed rotational band structure) for such nuclei having states dominated by identical nucleon configuration.  

For many particles in $j^n$ configuration, the states with different seniorities become orthogonal in the seniority scheme. On operating 
$S_j^+$ on the $J,v$ state in the $j^v$ configuration, one obtains the same $J,v$ state in the $j^{v+2}$ configuration. When we operate $S_j^-$ to annihilate a pair on the $J$ state in the $j^{v+2}$ configuration, and it has zero pairing energy, the seniority of that state becomes $v+2$. In a given $j^v$ configuration, two or more independent states are possible, both having the same $J$ and same seniority $v$. To differentiate them further, one needs to introduce an additional quantum number $\alpha$ for constructing an orthogonal basis of such states. Any two orthogonal $j^v$ configuration states remain to be orthogonal even after operating $S_j^+$ for $\frac{n-v}{2}$ times. Therefore, the physical understanding obtained for the states in the $j^v$ configuration can provide a common basis for the same states in any $j^n$ configuration. 

The single-particle electromagnetic transition operators are Hermitian in nature. The matrix elements of single-particle operators with odd-tensor disappear in the $j^n$ configuration, according to the Wigner--Eckart theorem. It is not possible to have odd-tensor operators with a $\kappa=0$ component of $s=1$ rank. Such operators exclusively behave as quasi-spin scalars with rank $s = 0$; that is, no distinct values of s are permitted for the states they connect. This can be represented as follows:
\begin{eqnarray}
\langle j^n v l J || T^{(k=odd)} || j^n v' l' J' \rangle & = \quad \quad \langle s, S_j^0 | T_{\kappa=0}^{s=0} | s', S_j^0 \rangle \nonumber \\
= & (-1)^{s-S_j^0} \langle s || T^{s=0} || s \rangle \begin{pmatrix}
s & 0 & s \\
-S_j^0 & 0 & S_j^0 \\  
\end{pmatrix}
\end{eqnarray}
where $s=s'$; this eventually requires $v=v'$, since $s=\frac{1}{2}(\Omega-v)$, $s'=\frac{1}{2}(\Omega-v')$  and $S_j^0=\frac{1}{2}{(n-\Omega)}$. The involved 3j-symbol becomes 1. A similar operator remains valid in the $j^v$ configuration and can be obtained by substituting \textit{n} to \textit{v}, i.e., $S_j^0=-s$. 
The non-vanishing matrix elements of single-particle odd tensor operators therefore become the particle number (\textit{n}) independent. This result can be written as follows:

\begin{equation}
\langle j^n v l J M | T_\kappa^{(k=odd)} | j^n v' l' J' M' \rangle = \langle j^v v l J M | T_\kappa^{(k=odd)} | j^v v' l' J' M' \rangle \hspace{1mm} \delta_{v,v'}
\end{equation} 
 
Such odd tensor single-particle operators only connect the states of the same seniority \textit{v}; i.e., \textit{v} remains preserved. Due to the way that magnetic operators operate in a single-j shell, we may infer the behavior of magnetic transitions and magnetic dipole moments from this. According to this, identical nucleons in a single-j shell exhibit magnetic moments that behave independently of particle number, and the matrix elements of magnetic dipole moments in the $j^n$ configuration can be converted to the $j^v$ configuration with coefficient 1. To determine the magnetic moment of the same \textit{J}, \textit{v} state in different isotopes or isotones in the single-j shell, one only needs to know the magnetic moment for the seniority \textit{v} state in question. The magnetic moment $(\mu)$ of the state with a total angular momentum \textit{J} arising from \textit{n} identical nucleons in a single-j shell can be written as

\begin{equation}
\vec{\mu}=\sum_{i=1}^n g \vec{j_i} =g \sum_{i=1}^n \vec{j_i} = g\vec{J}
\end{equation}

The magnetic moment $\mu/J$ ratio determines the g-factor. For every seniority state resulting from the $j^n$ configuration, it stays largely unchanged. The seniority suggests that the particle-number variation of the g-factor is constant. This means that the g-factor of any seniority state (\textit{v} =even in even-A or \textit{v} = odd in odd-A) in the $j^n$ configuration will be about equivalent to the g-factor of a single seniority (\textit{v} = 1) state.

The even-tensor operators, on the other hand, turn out to be the $\kappa=0$ component of the quasi-spin vectors $(s=1)$. Using the Wigner--Eckart theorem, the change in \textit{s} values of the states connected by the even-tensor operators is now allowed at most by 1. The corresponding matrix elements between the states having a respective seniority of \textit{v} and \textit{v}' would be non-vanishing only if their seniorities are equal to $(\Delta v=0)$ or differ at most by 2 $(\Delta v=2)$. In the quasi-spin space for the $j^n$ configuration, this can be written as:

\begin{eqnarray}
\langle j^n v l J || T^{(k=even)} || j^n v' l' J' \rangle &= \quad \quad
\langle s, S_j^0 | T_{\kappa=0}^{s=1} | s', S_j^0 \rangle \nonumber \\ =& (-1)^{s-S_j^0} (s||T^{s=1}||s') \begin{pmatrix} s & 1 & s' \\ -S_j^0 & 0 & S_j^0 \end{pmatrix}
\end{eqnarray} 

Such matrix elements' \textit{n}-dependence results from the related 3j-symbol and differs depending on whether \textit{s}' =\textit{s} and \textit{s}' =s $\pm 1$. To start with, when \textit{s}' =\textit{s}, i.e., \textit{v}' = \textit{v}, the quasi-spin matrix elements can be given as

\begin{eqnarray}
\langle s, S_j^0 | T_{\kappa=0}^{s=1} | s, S_j^0 \rangle &= \frac{2S_j^0}{\sqrt{2s(2s+1)(2s+2)}} (s||T^{s=1}||s)  \nonumber \\
 &= \frac{(n-\Omega)}{ \sqrt{2s(2s+1)(2s+2)}} (s||T^{s=1}||s)  \label{quasispin}
\end{eqnarray} 

A similar relation holds true for the $j^v$ configuration and can be simply obtained by replacing \textit{n} with \textit{v} in Equation (\ref{quasispin}). In electromagnetic transitions, the angular component is typically expressed in terms of spherical harmonics $Y^L$. The corresponding matrix elements between the states with initial and final parities of \textit{l} and \textit{l}' in the $j^n$ configuration would reduce to $j^v$ configuration: 

\begin{equation}
\langle j^n v l J M | Y_\kappa^L | j^n v l' J' M'\rangle =\Bigg[ \frac{\Omega-n}{\Omega-v} \Bigg] \langle j^v v l J M | Y_\kappa^L | j^v v l' J' M'\rangle \label{dv0}
\end{equation} 

which is true for nonzero even \textit{L} values, $L>0$, where \textit{L} is the allowed multipole value and represents the angular-momentum transfer. For the \textit{n}-particle $j^n$ and \textit{n}-hole $j^{2j+1-n}$ configurations, these matrix elements of electric even tensor operators $(L > 0)$ have opposing signs but are identical in magnitude. The matrix component reverses its sign after becoming zero at the middle of the shell, where $n=\frac{2j+1}{2}=\Omega$. The condition of a hindered transition is therefore supported at the middle, and so is the seniority isomerism. In any scheme, $Y_0^{(0)}$ with \textit{L} = 0 has only diagonal components. All of these components are proportionate to the number operator $n=\sum_m a_{jm}^+ a_{jm}$ and equal, i.e., \linebreak $Y_0^{(0)}=\sum_{m,m'} \langle jmjm'|jj00 \rangle a_{jm}^+ \tilde{a}_{jm'}=\frac{1}{\sqrt{2j+1}} \sum_m a_{jm}^+ a_{jm}$.

Similarly, when $s'=s\pm1$, i.e. $v'=v\mp2$,

\begin{eqnarray}
\langle j^n v l J M | Y_\kappa^L | j^n v\mp2, l' J' M'\rangle =\Bigg[ \sqrt{\frac{(n-v+2)(2\Omega+2-n-v)}{2(2\Omega+2-2v)}} \Bigg] \\ \nonumber
\times \langle j^v v l J M | Y_\kappa^L | j^v v\mp2, l' J' M'\rangle \quad \text{($L>0$, even)} \label{dv2}
\end{eqnarray} 

\begin{figure}[!ht]
\includegraphics[width=0.7\textwidth]{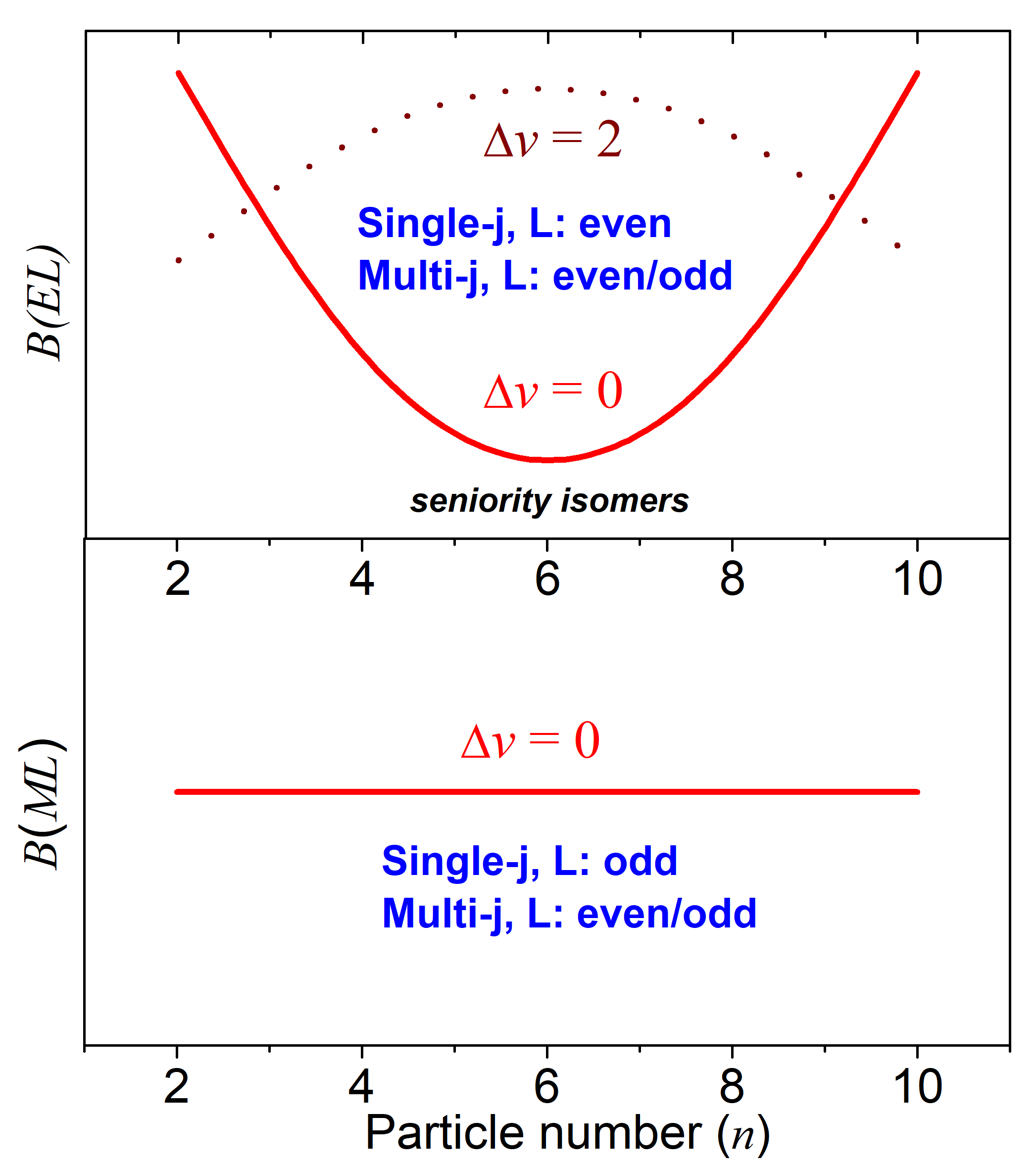}
\caption{\label{fig:transition}(Color online) Schematic variation of electromagnetic reduced transition probabilities in both single-j shell and multi-j shell by using the electromagnetic and (generalized) seniority selection rules. The additional possibility of odd electric tensor decaying seniority isomers has been pointed out in the upper panel.}
\end{figure}

Therefore, for a single-j shell, the related electric transitions for the even rank tensors have $\Delta v=0$, or $2$. In a single-j $j^n$ configuration, the reduced electric transition probabilities $B(EL)$ between $J_i$ and $J_f$ states can be written as,
 
\begin{equation}
B(EL)=\frac{1}{2J_i+1}|\langle j^n v l J_f || \sum_i r_i^L Y^{L}(\theta_i,\phi_i) || j^n v' l' J_i \rangle |^2
\end{equation}
where the total pair degeneracy $\Omega= \frac{1}{2}(2 j +1)$ and only even values of $l, l'$ and \textit{L} are allowed. The seniority reduction formulae for the seniority conserving $\Delta v=0$ transitions in Equation (\ref{dv0}) and seniority changing $\Delta v=2$ transitions in Equation (\ref{dv2}) simplify a \textit{n}-body problem to a \textit{v}-body problem. These relations imply a parabolic behavior for $B(EL)$, as shown in Figure~\ref{fig:transition}.

The idea of seniority in a single-j shell should therefore be generalized to a multi-j environment because the actual situation in nuclei may entail multiple closely spaced orbitals in their active nucleonic valence space. The most straightforward method entails the extension of the multi-j scheme's quasi-spin scheme, which has been around for a while and is typically credited to Kerman~\cite{kerman1961, kerman19611} and Helmers~\cite{helmers1961}. The idea of generalized seniority in a multi-j shell for diverse degenerate orbitals was first suggested by Arima and Ichimura~\cite{arima1966}. This concept was expanded by Talmi to include the realistic scenario of several non-degenerate orbitals~\cite{talmi1971,shlomo1972}. A seniority scheme was also proposed by Otsuka and Arima~\cite{otsuka1978} for a variety of non-degenerate single-particle orbitals, which included particle number conservation along with the BCS approximation and SU(2) quasi-spin as two extreme limits. In the simplest case of many degenerate orbitals~\cite{arvieu1966}, the concept of seniority can be extended to generalized seniority by defining the multi-j pair creation operator as~\cite{talmi1993}:

\begin{equation}
S^{+} = \sum_{j} S_{j}^{+},
\end{equation} 
where the summation takes care of all the active valence orbitals. This operator effectively sums up the operators for each individual orbital from the active multi-j space. The operators, $S^+$, $S^-$ (Hermitian conjugate of $S^+$) and $S^0$ $(S^0=\frac{1}{2}{(n-\Omega)}$,  $n=\sum_{j} {n_j} $, \linebreak $\Omega=\frac{1}{2} \sum_{j} (2j+1) $), also behave as the generators of the $SU(2)$ Lie group. Thus, a seniority scheme for identical nucleons in a multi-j shell of multiple degenerate orbitals is defined by the system of eigen states of $S^2$, or those of $2 S^+ S^-$. It incorporates all the characteristics of the single-j seniority scheme that come from the SU(2) algebra and is known as the generalised seniority scheme. The pairing interaction $-2G S^+ S^- = -2G (S^2 - S^0(S^0-1))$, which resembles the single-j relation, has energy eigenvalues that are given by

\begin{equation}
-G(2s(s+1)-\frac{1}{2} {(\Omega-n)(\Omega+2-n)}) = -G \Bigg(\frac{n-v}{2} {(2\Omega+2-n-v)}\Bigg)  
\end{equation}

\vspace{-6pt}

The total quasi-spin can be obtained as $s=\frac{1}{2}(\Omega-v)$, and $S^0=\frac{1}{2}{(n-\Omega)}$ with $v$ being the generalized seniority. The total number of particles in a multi-j shell is equal to the sum of the occupancies of all active j-orbits, or $n=\sum_{j} {n_j} $, and the corresponding pair degeneracy is equal to $\Omega=\frac{1}{2} \sum_{j} (2j+1) $. Similar to any quasi-particle model allowing all conceivable configuration mixings and shared occupancy among active orbitals, such generalized seniority states support all possible configuration mixings. For instance, ${(S^+)}^n |0> $, where $n=\sum_j n_j$ and $S^{+} = \sum_{j} S_{j}^{+}$, may be used to generate the generalized seniority $v = 0$ state. In order to create a generalized seniority $v = 0$, ${(S^+)}^n |0> $ state, each state with ${(S_j^+)}^{n_j} |0> $ is now combined with various possible single-j configurations. 

In a multi-j shell, a single-particle tensor operator with rank $k$ and component $\kappa$, $T_\kappa^{(k)}$ follows the commutation $[S^+, T_\kappa^{(k)}] = 0$ for odd \textit{k} values, and as a result, the odd Hermitian tensors behave as quasi-spin scalars. The quasi-spin vector's $k= 0$ component is how the even-rank single-particle Hermitian tensor operators behave. These findings are directly relevant to electromagnetic transitions where the spherical harmonic operator is Hermitian in nature. These findings are a straightforward generalization of the single-j seniority. To handle multi-j shells with orbitals of different parities, the pair-creation operator can be redefined as~\cite{maheshwari2016}

\begin{equation}
S_1^{+} = \sum_{j} (-1)^{l_j} S_{j}^{+}
\end{equation} 
where the phase factor $(-1)^{l_j}$ involves $l_j$ to take care of the parity of each orbital. With spherical harmonics $Y^L$, the total phase factor for the electromagnetic transitions is changed to $(-1)^{l+l'+L}$ ~\cite{maheshwari2016}. Here, \textit{l}, \textit{l}' denote the parities of the initial and final states connected by the electromagnetic transition, and \textit{L} specifies the multipole of the relevant electric/magnetic operator. For the electromagnetic transitions in the multi-j shell, the spherical harmonic tensor, $Y^L$, behaves as: 

\begin{itemize}
\item A quasi-spin scalar if $l+l'+L$ is odd.
\item $\kappa=0$ component of the quasi-spin vector if $l+l'+L$ is even.
\end{itemize}   

Regardless of the nature of the $L$ value owing to the selection criteria for electromagnetic transitions, the sum $l+l'+L$ always remains even for electric transitions. Similarly, regardless of the nature of the $L$ value, the sum $l+l'+L$ is always odd for magnetic transitions \cite{maheshwari2016}. Consequently, the magnetic transitions irrespective of the involved nature of the tensor would behave as seniority-preserving transitions. The magnetic moments (i.e., g-factors) would be nearly constant with respect to nucleon number for a given state and configuration. On the other hand, the electric transitions would support a parabolic behavior on varying the nucleon number for both odd and even multipole tensors. \linebreak The basic multi-j quasi-spin algebra was known since the 1960s, although no one pointed out the crucial consequences of such generalized seniority selection rules on the electromagnetic selection rules until 2016~\cite{maheshwari2016}. This has opened up the prospect of having good seniority states with $M2,M4...,$ or $E1, E3...$ transitions, as opposed to the widely held notion that good seniority states can only be obtained with $M1,M3..,$ or $E2, E4..$ transitions. 
The choice of phase factor $(-1)^{l_j}$ by Arvieu and Moszokowski~\cite{arvieu1966} turned out to be crucial to guide these electromagnetic selection rules and is a consequence of the surface--delta interaction for which the generalized seniority is an approximate quantum number. This phase factor hence remains to be the most appropriate for the excellent description of generalized seniority states \cite{maheshwari2016,maheshwari20161,kota2017,jain2017,jain20171}. If the phase choice is different, then the corresponding electromagnetic selection rules would also be different, as shown by Kota~\cite{kota2017}. Thus, a key component of the realistic nuclear interaction is the pairing interaction. The discussed generalized seniority in this overview is different from Talmi’s prescription of generalized seniority with a pair-creation operator of $\sum_j \alpha_j S_j^+$, where $\alpha_j$ can take unequal and non-integer values. Here, the $\alpha_j$ coefficients are equal apart from a change of sign decided by the orbital angular momentum of the given orbital $l_j$.

\section{Seniority (and Generalized Seniority) Isomers: Where and Why?}\label{sec3}

In a single-j shell, magnetic transitions are allowed with odd-multipole tensors while the electric transitions are only allowed with even-multipole tensors due to electromagnetic transition selection rules. The seniority selection rules dictate that the odd-tensor magnetic operators preserve the seniority and lead to a particle number-independent variation of the matrix elements. Therefore, the magnetic moments---that is, g-factors for a given $J,v$ state in $j^n$ configuration---have nearly constant value on changing the nucleon number. The seniority selection rules allow the even-tensor operators to either preserve the seniority $\Delta v=0$ or change it by 2, $\Delta v=\pm 2$. As a result, the reduced transition probabilities $B(EL)$ for electric transitions have a parabolic behavior. For $\Delta v=0$ transitions, the $B(EL)$ parabola has a dip in the middle of the single-j shell, whereas for $\Delta v=2$ transitions, it has a peak in the middle. The minimum value of $B(EL)$ ($L>0$, even) infers extra hindrance at the middle and gives rise to the isomerism due to seniority selection rules. These isomers occur due to particle-hole symmetry cancelling the electric matrix elements at the middle for good seniority states. Such a condition is fulfilled for the electric quadrupole (even multipole) transitions in a single-j shell, and therefore, \textit{E}2 seniority isomers are well known, particularly for semi-magic nuclei having identical nucleons. The elegance of the seniority scheme lies in the fact of reducing an \textit{n}-body problem to the \textit{v}-body problem. The seniority reduction equations can be used to determine how matrix elements change as the nucleon number \textit{n} changes if the solution of matrix elements in the $j^v$ configuration is known. Even-multipole magnetic and odd-multipole electric transitions are prohibited, because the single-j shell cannot undergo a parity change.

In the multi-j shell with various orbitals of different parities, the generalized seniority and electromagnetic selection rules altogether allow the parity mixing, and so are the even/odd-electric as well as even/odd-magnetic transitions \cite{maheshwari2016}. As we transition to the multi-j shell, the picture of seniority isomerism alters significantly due to these subtle but significant differences. As seen in Figure \ref{fig:transition}, the magnetic transitions now exhibit particle number-independent behavior and preserve seniority for both even and odd multipole tensors. Contrarily, the electric transitions behave as the $\kappa=0$ component of the quasi-spin vector regardless of whether \textit{L} is even or odd. Therefore, for both even and odd multipole tensors, the reduced electric transition probabilities between the identical seniority states exhibit a parabolic minimum in the middle of the active shell, whereas seniority changing transitions exhibit a maximum at the middle, as seen in Figure \ref{fig:transition}. Accordingly, it follows that one should obtain greater half-lives for the even and odd tensor seniority preserving electric transitions at the middle of the multi-j shell. Due to this straightforward finding, a novel class of seniority isomers that decay via odd-electric transitions has been identified~\cite{maheshwari2016}. 

For a multi-j configuration $\tilde{j} = j \otimes j'....$ specified using the total pair degeneracy $\Omega= \frac{1}{2}(2 \tilde{j} +1)= \frac{1}{2} \sum \limits_j (2j+1)$, the reduced transition probabilities $B(EL)$ between initial $J_i$ and final $J_f$ states are given by,
 
\begin{equation}
B(EL)=\frac{1}{2J_i+1}|\langle \tilde{j}^n v l J_f || \sum_i r_i^L Y^{L}(\theta_i,\phi_i) || \tilde{j}^n v' l' J_i \rangle |^2 \label{be2}
\end{equation}

Inferring from this is that regardless of the \textit{L} (even or odd) values, the $B(EL)$ values display a parabolic behavior in the multi-j case. For the reduced electric matrix elements with $\Delta v=0$ and $\Delta v=2$ transitions, the generalized seniority reduction formulas may be stated as follows

\begin{equation}
\langle \tilde{j}^n v l J_f ||\sum_i r_i^L Y^{L}(\theta_i,\phi_i)|| \tilde{j}^n v l' J_i \rangle = \Bigg[ \frac{\Omega-n}{\Omega-v} \Bigg] \langle \tilde{j}^v v l J_f ||\sum_i r_i^L Y^{L}(\theta_i,\phi_i)|| \tilde{j}^v v l' J_i \rangle  \label{be2_dv0}
\end{equation}

\begin{eqnarray}
\langle \tilde{j}^n v l J_f ||\sum_i r_i^L Y^{L}(\theta_i,\phi_i)||\tilde{j}^n v\pm 2, l' J_i \rangle  &=& \Bigg[ \sqrt{\frac{(n-v+2)(2\Omega+2-n-v)}{4(\Omega+1-v)}} \Bigg] \nonumber\\ && \langle \tilde{j}^v v l J_f ||\sum_i r_i^L Y^{L}(\theta_i,\phi_i)|| \tilde{j}^v v\pm 2, l' J_i \rangle  \label{be2_dv2}
\end{eqnarray}

The Q-moments in even--even nuclei change signs in the mid shell. The generalization to a  multi-j shell suggests that such moments should have opposite signs at the beginning and the end of major shells. The realistic trends are, however, complicated by the multi-j shell; this qualitative feature is a well-known empirical characteristic of heavy nuclei, and it contrasts markedly with that for odd-tensor operators that are independent of $n$~\cite{casten}. 

The magnetic dipole moment for a given multi-j configuration $\tilde{j}$ may simply be written as:
\begin{equation}
\langle \tilde{j}^n |\hat{\mu} | \tilde{j}^n \rangle = \langle \tilde{j}^v |\hat{\mu}| \tilde{j}^v  \rangle \label{magmoment}
\end{equation} 
where $\tilde{j}=j \otimes j'....$ represents a multi-j valence shell with a total occupancy of \textit{n} shared among active orbitals. It may be noted that the generalized seniority Equations (\ref{be2_dv0})--(\ref{magmoment}) can directly be applied when the multiplicity $L$ takes same values for the given transition in the isotopic/isotonic chain. The magnetic moment of identical nucleons in the mixed-j $\tilde{j}^n$ configuration is given by~\cite{maheshwari2019}

\vspace{-6pt}

\begin{equation}
\vec{\mu}=g \sum_{i=1}^n {\vec{\tilde{j}}_i} = g\vec{J}
\end{equation}

As a result, as with the single-j case, the g-factors for the reasonably good generalized seniority states in a multi-j configuration also display a nearly constant and particle number independent behavior. It would be beneficial to combine the notion of generalized seniority with the well-known Schmidt model, known as the Generalized Seniority Schmidt Model (GSSM) \cite{maheshwari2019}, in order to acquire the values numerically (without needing to calculate magnetic reduced matrix elements). Schmidt equations \cite{schmidt1937} are often used to determine the g-factors in odd-A nuclei, especially those close to the shell closures. On defining the similar expressions in GSSM phenomenologically, by using the multi-j configuration $\tilde{j}=j \otimes j'....$ corresponding to the total pair degeneracy $\Omega={\frac{1}{2} (2\tilde{j}+1)}=\frac{1}{2} \sum_j (2j+1) $ for any given generalized seniority \textit{v} state, one obtains \cite{maheshwari2019}:
\begin{eqnarray}
g  =& \frac{1}{\tilde{j}} \Bigg[ {\frac{1}{2} g_s}+ (\tilde{j}- \frac{1}{2}) g_l \Bigg]; \tilde{j}=\tilde{l}+\frac{1}{2}\nonumber\\ 
   =& \frac{1}{\tilde{j}+1} \Bigg[ -\frac{1}{2} g_s + (\tilde{j}+ \frac{3}{2}) g_l \Bigg]; \tilde{j}=\tilde{l}-\frac{1}{2} 
\end{eqnarray}
where $g_s=5.59$ n.m. and $g_l=1$ n.m. for protons, while $g_s=-3.83$ n.m. and $g_l=0$ n.m. for neutrons. All seniority states derived from identical nucleons in a multi-j shell must have g-factor values that are substantially equivalent to the g-factor of a single nucleon (seniority \textit{v} = 1) in the same multi-j shell, provided the amount of configuration mixing is almost similar. This is also a signature of nearly good generalized seniority, and the resulting effective interaction should be approximately diagonal in generalized seniority.

Similar to good seniority states originating in a single-j shell, good generalized seniority states' excitation energies are predicted to exhibit valence particle number independence. The two-body matrix elements for the 0\textsuperscript{+} (fully coupled) state in the multi-j scenario for a two-body odd-tensor interaction $V_{ik}$ may be defined as $V_0= \langle {\tilde{j}}^2 J=0 |V_{ik} | {\tilde{j}}^2 J=0 \rangle $. In terms of a $j^v$ configuration, the matrix element for a $j^n$ configuration may be represented as~\cite{maheshwari20191}:

\begin{equation}
\langle {\tilde{j}}^n v l J | \sum_{i<k}^n V_{ik} | {\tilde{j}}^n v l' J \rangle = \langle {\tilde{j}}^v v l J |  \sum_{i<k}^n V_{ik} | {\tilde{j}}^v v l' J \rangle  + \frac{n-v}{2} V_0 \delta_{l,l'}
\end{equation}

If \textit{v} is equal to 0 or 1, the first term in the equation above becomes 0. The ground states of even--even nuclei (\textit{v} = 0) and even--odd/odd--even nuclei (\textit{v} = 1) may thus be readily calculated as $\frac{n}{2}V_0$ and $(\frac{n-1}{2})V_0$, respectively. It suggests that the number of pairs of particles connected to the \textit{J} = 0 state, where the total number of particles is $n=\sum_j n_j$, determines the ground state energy. Therefore, the energy difference in even--even nuclei between the generalized seniority $v=2, J \ne 0$ and $v=0, J = 0$ (ground) states may be expressed as follows: 

\vspace{-6pt}

\begin{equation}
E({\tilde{j}}^n, v=2, J)- E({\tilde{j}}^n, v=0, J=0) = \text{constant} 
\end{equation}

As a result, for a given multi-j configuration, the energy difference is independent of the valence particle number. For example, throughout the isotopic chain, the first excited $2^+$ states in Sn isotopes are detected at practically constant energy~\cite{maheshwari20191}. This is also true for even--even Sn isotopes with high-spin ${10}^+$ isomers~\cite{jain2017}. The ${27/2}^-$ isomers in the nearby odd-A Sn isotopes~\cite{jain2017} also behave in a similar manner.

\section{Seniority Isomers in Various Mass Regions}\label{sec4}

Due to particle--hole symmetry, the reduced matrix components for electric transitions vanish in the middle for transitions between the same seniority states, allowing for the existence of seniority isomers. In the following, we discuss the isomeric examples for the validity of seniority and generalized seniority schemes, where seniority is a reasonably pure quantum number. Due to the involved seniority and symmetries, the seniority isomers behave very similar to each other despite the different valence spaces and orbitals. However, the occurrence of such isomers is not straightforward when the dominating orbital is surrounded by high-j orbitals, as is the case of heavier mass nuclei. It may be noted that a part of the following discussion for the seniority isomers, particularly due to dominating intruder orbitals in Sn, Pb isotopes and $N=82$ isotones can also be found in Ref. \cite{jain2021}. \linebreak We hereby extend the discussion to cover the present status of seniority isomers by surveying each possible isotopic/isotonic chain having nuclei with either protons or neutrons at the closed shell and the nuclei with two-particles/holes away from the respective closed shell. 

\subsection{Ca Isotopes}

Ca isotopes are known to be the starting examples for good seniority. The lowest orbital beyond the $^{40}$Ca core is the $f_{7/2}$ orbital, and the ground states of $^{41}$Ca to $^{48}$Ca are formed by adding neutrons to the $f_{7/2}$ orbital. The most-aligned $v=2$ state in the $f_{7/2}^2$ configuration would be the $6^+$ state. For two $f_{7/2}$ particles/holes in $^{42}$Ca and $^{46}$Ca, the $6^+ (v=2)$ isomers are experimentally known with respective half-lives of 5.28 \textit{ns} \cite{marmor1970}, and 10.4 \textit{ns} \cite{kutschera1975}. The known g-factor value of $-0.425$ n.m. for this $6^+$ isomer in $^{42}$Ca also confirms the purity of the configuration. The seniority and configuration predict almost the same g-factor for the $6^+$ isomer in $^{46}$Ca with a negative sign.

However, the mid-shell nucleus $^{44}$Ca with four particles in $f_{7/2}$ does not support a longer-lived $6^+$ isomer as per the seniority expectations. There would be two $4^+$ states for the four particles in $f_{7/2}$: one with $v=2$ and another with $v=4$. Due to the Berry phase situation in the middle of the shell \cite{dobon2021} for $^{44}$Ca, the allowed $\Delta v=4$ interaction results in the lower lying $v=4, J=4$ state providing an additional and faster decay branch for the $6^+$ state using a $\Delta v=2$ transition and thus a no longer-lived $6^+$ state.    

\subsection{Ni Isotopes}

There are no seniority isomers in $^{58-66}$Ni isotopes, since there is no high-j orbital involved. However, the activeness of the $g_{9/2}$ orbital has led to interest in searching the $8^+$ seniority isomers for neutron-rich Ni isotopes beyond the $^{68}$Ni core \cite{lewitowicz1999, rajabali2014, soderstrom2015}. The first orbital for which seniority-mixing effects may be anticipated is the $g_{9/2}$ orbital of identical particles. As expected, the $8^+$ isomers are known to exist for two $g_{9/2}$ particles/holes configuration in $^{70,76}$Ni with respective half-lives of 0.232 $\mu s$ and $0.547$ $\mu s$ \cite{atlas22}. The two neutron g$_{9/2}^2$ particles' based excited states $J = 2^+, 4^+, 6^+$ and $8^+$ are seen in a regular seniority-like band for the spectra of $^{70}$Ni \cite{lewitowicz1999}. Additionally, $^{76}$Ni possesses two neutron g$_{9/2}^{-2}$ holes based on the yrast sequence of excited states with $J = 2^+, 4^+, 6^+$ and $8^+$~\cite{soderstrom2015}. Two particle--hole symmetry therefore prevails. However, there is an additional intruder 2\textsuperscript{+} state in the case of $^{70}$Ni but not in $^{76}$Ni \cite{isacker2011}. 

The explanation behind the absence of seniority 8\textsuperscript{+} isomers for four $g_{9/2}$ particles/holes in $^{72}$Ni and $^{74}$Ni~\cite{sawicka2003} has been provided using seniority mixing arguments \cite{isacker2011}. As soon as four particles/holes start to occupy the $g_{9/2}$ orbital, the possible states would be $0^+$ (twice), $2^+$ (twice), $3^+$, $4^+$ (thrice), $5^+$, $6^+$ (thrice), and $7^+$, $8^+$, $9^+$, ${10}^+$, ${12}^+$ \cite{mayer1955}. 
The $8^+$ states in both of these Ni isotopes are able to decay by the allowed and faster $\Delta v=2$ transition to the second $6^+$ state, which has $v = 4$, since two levels for $J= 4^+$ and $J=6^+$ exist and are fairly near in energy. According to Equation (\ref{dv2}), the enhanced transition probability between the $v=2,8^+$ and $v=4,6^+$ states causes the 8\textsuperscript{+} state's half-life to be shorter, resulting in no isomer. For these 8\textsuperscript{+}  seniority isomers (states) in neutron-rich Ni isotopes, there are no moment measurements available.

\subsection{Sn Isotopes}

\begin{figure}[!ht]
\includegraphics[width=0.85\textwidth]{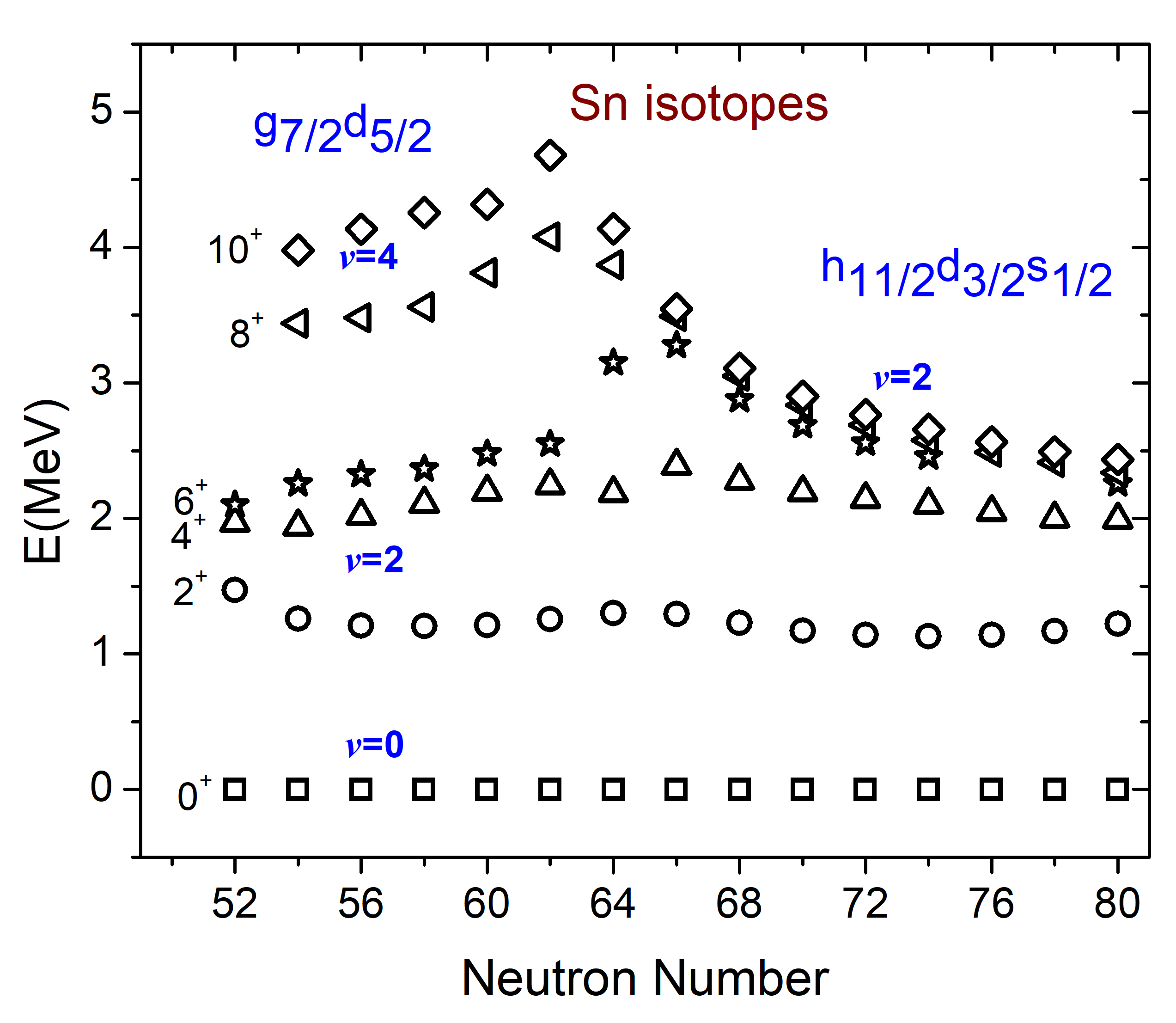}
\caption{\label{fig:snevenenergy}(Color online) {Empirical energy} 
systematics for the 0\textsuperscript{+} to 10\textsuperscript{+} states in Sn isotopes from \textit{N} = 52 to 80, where \textit{v} is the seniority quantum number. The major contributing orbitals are also pointed out before and after the middle of the neutron 50--82 valence space.}
\end{figure}

The first $2^+$ states in semi-magic nuclei have a nearly particle number independent energy variation, which has long been recognized~\cite{talmi1993}. The Sn isotope chain, which extends from the doubly magic $^{100}$Sn to the doubly magic $^{132}$Sn and beyond~\cite{astier2012,iskra2014,simpson2014,iskra2016}, is the traditional place to illustrate seniority. Sn exists in 41 different known isotopes, ranging from $^{99}$Sn to $^{139}$Sn. There are just seven stable ones that are found in nature. This puts these isotopes at the forefront of nuclear physics from a theoretical and experimental perspective. The $0^+$ ground states in even--even Sn isotopes are fully pair-correlated states with $v = 0$. The first $2^+, 4^+$ and $6^+$ states are $v=2$ states, resulting in an approximately constant energy variation in $N = 50-82$ valence space; nonetheless, the impact of various orbitals is apparent in their configurations on varying the neutron number. A single-j seniority model cannot account for this behavior throughout the whole $N = 50-82$ valence space. \linebreak The dominating orbitals before the middle (with effective $\tilde{j}=19/2$) and after the middle (with effective $\tilde{j}=23/2$) change and so does the energy difference between the $0^+$ and $2^+$ states from $\approx 1.2$ MeV to $\approx 1.1$ MeV~\cite{maheshwari20191}. While the $8^+$ and ${10}^+$ states may be inferred as $v = 2$ states in heavier Sn isotopes where the $h_{11/2}$ orbital starts to predominate, the $8^+$ and ${10}^+$ states appear to be $v = 4$ states in the region of influence of the $g_{7/2}$ and $d_{5/2}$ orbitals near $^{100}$Sn. This also leads to the finding that the energy gap between the identical generalized seniority states $(\Delta v=0)$ in even--even nuclei is almost constant. With the exception of the region near $N = 64$, the energy difference, for instance, between $2^+$ and $4^+$ and between $4^+$ and $6^+$ is almost constant; see Figure \ref{fig:snevenenergy}. The ${10}^+$ isomers decay to the $8^+$ states by nearly constant $E2$ gammas for $N>64$ even--even Sn isotopes~\cite{jain2017}. These observations can also be extended to odd-A nuclei having generalized seniority changing transitions, where the $v=3$, ${27/2}^-$ isomers, decaying via nearly constant $E2$ gammas to the lower-lying $v=3$, ${23/2}^-$ states, also support a constant energy gap with respect to the $v=1$, ${11/2}^-$ states~\cite{jain2017}. 

\begin{figure}[!ht]
\includegraphics[width=0.9\textwidth]{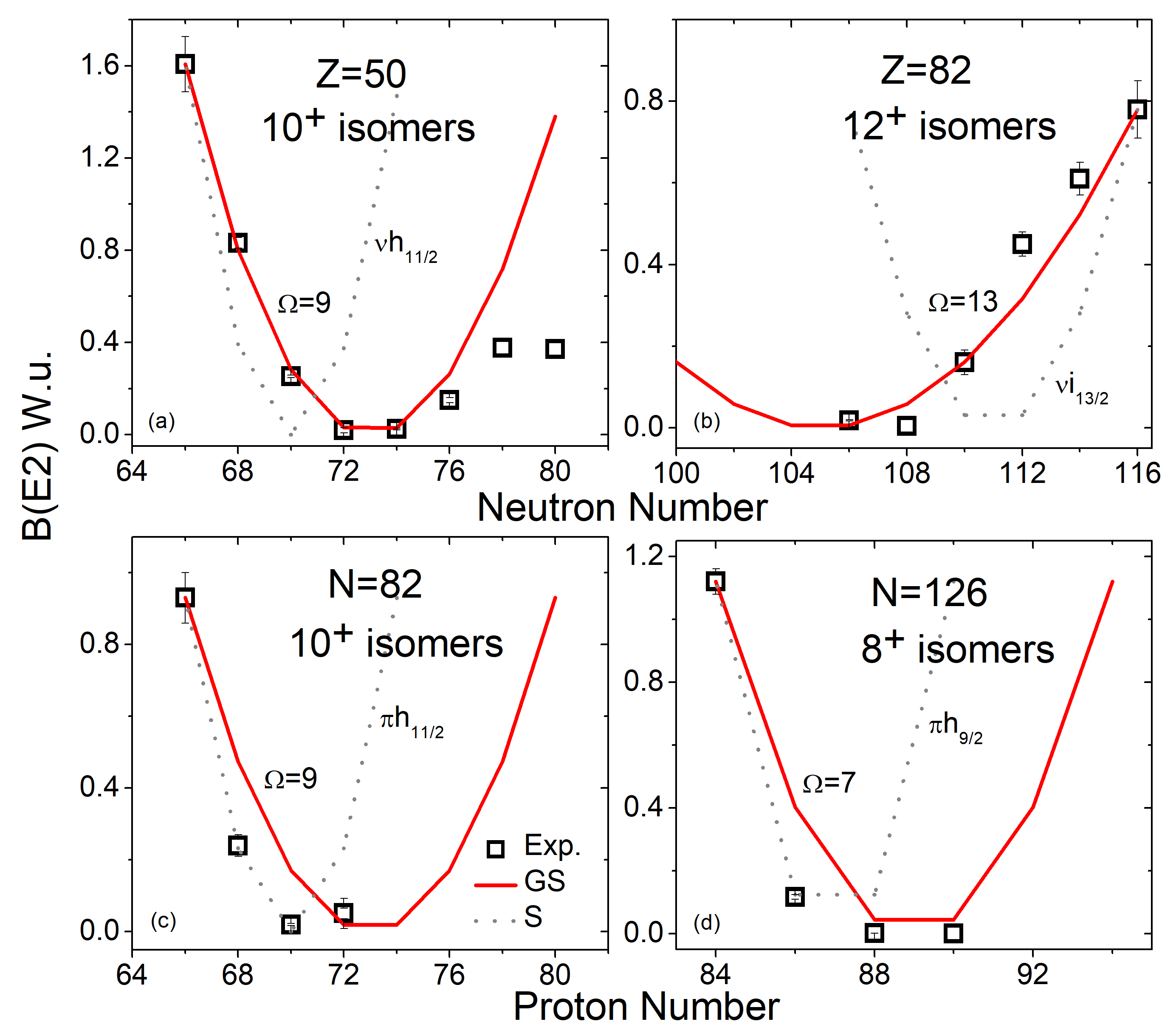}
\caption{\label{fig:BE2-GS}(Color online) {B(E2) variation} 
of \textit{v} = 2 isomers in various semi-magic chains; (a) for $Z=50$ isotopes, (b) for $Z=82$ isotopes, (c) for $N=82$ isotones, and (d) for $N=126$ isotones. GS and S refer to the generalized seniority and seniority results, respectively. The uncertainties in the experimental data~\cite{ensdf} are shown but mostly lie within the size of symbol.}
\end{figure}

It should be observed that the ${10}^+$ and ${27/2}^-$ states for $N > 64$, Sn isotopes stay very close in measured excitation energies \cite{daly1980,fogelberg1981,daly1986,lunardi1987,broda1992,mayer1994,daly1995,pinston2000,zhang2000,lozeva2008} because they undergo the same structural transition from $0^+$ to ${10}^+$ states in even--even isotopes to that from ${11/2}^-$ to ${27/2}^-$ states in odd-A isotopes. The ${10}^+$ and ${27/2}^-$ isomeric states have been adopted as seniority $v=2$ and $v=3$ states, respectively, generating only from $h_{11/2}$ neutrons using seniority arguments. Although the single-j seniority scheme suggests the occurrence of $E2$ decaying isomers, a full explanation for the existence of these isomeric states beyond the occupancy of the $h_{11/2}$ orbital confirms the need for a multi-j mixed configuration. The isomeric half-lives also display a maximum at $N=73$, which is the signature of the half-filled $h_{11/2}$ orbital. This clearly hints that for $N>64$, the g$_{7/2}$ and d$_{5/2}$ orbitals are almost filled up and can be considered to be frozen, bringing the neutron core from $N=50$ to $N=64$. The remaining valence space can accommodate 18 neutrons consisting of $h_{11/2}$, $d_{3/2}$, and $s_{1/2}$ with the middle at $N=73$. The corresponding transition probability parabola touches the bottom at $N=73$ as per the generalized seniority shown in Figure \ref{fig:BE2-GS}a. The half-lives, therefore, display a maximum near $N=73$ for these ${10}^+$ isomers~\cite{maheshwari2016}. These generalized seniority computations are carried out at the expense of fixing the reduced electric matrix elements using one of the experimental data. By limiting the focus to the first ${10}^+$ isomeric states (which are not a part of any rotational structure \cite{fotiades2011}), the deformed collective states that are assigned a two particle--two hole configuration \cite{bron1979,poelqeest1980,harada1988,savelius1998,gableske2001,wolinska2005,wang2010} in the even--even light mass Sn isotopes with $A = 110-118$ have been avoided for a systematic discussion. No indication of the disappearance of ${10}^+, v=2$ seniority isomers in the Sn isotopes also indicates toward the purity of the $8^+, v=2$ state in Sn isotopes. These $8^+$ states are also longer-lived excited states: that is, isomers for the heavier Sn isotopes. Similarly, the most-aligned $h_{11/2}^3,{27/2}^-$ isomers in odd-A Sn isotopes (with $N>64$) can be understood as arising from the $v=3$ configuration in multi-j $h_{11/2}\otimes d_{3/2}\otimes s_{1/2}$~\cite{jain2017}. The other $v=3,$ ${19/2}^+,{23/2}^+$ isomers with significant mixing from $d_{3/2}\otimes s_{1/2}$ orbitals are also known in these odd-A Sn isotopes~\cite{atlas22}. 

The identification of a higher seniority $v=4$, ${15}^-$ isomer in Sn isotopes \cite{pietri2011,astier2013,astier2012,iskra2014} has been of great interest due to its potential to probe effective interactions in many-body problems. The generalized seniority scheme, with a  multi-j configuration $h_{11/2} \otimes d_{3/2} \otimes s_{1/2} $ (pair degeneracy of $\Omega=9$)~\cite{maheshwari2016}, could explain the measured B(E1) rates for the $v=4, {13}^-$ isomers, and the B(E2) rates for the $v=4, {15}^-$ isomers in the Sn isotopes~\cite{astier2012,iskra2014}. Both the B(E2) and B(E1) rates support a parabolic trend for these neutron-rich Sn isotopes, irrespective of the odd or even nature of the involved electric tensor~\cite{maheshwari2016}. Any realistic effective interaction for describing these states should be nearly diagonal in the generalized seniority scheme. Although \textit{E}1 transition matrix elements require a mixing of lower-lying orbitals (say $g_{9/2}$) for the neutrons in 50--82 valence space, the single-point fitting (to the measurement) takes care of the required core excitations to generate weak \textit{E}1 transitions for the $v=4, {13}^-$ Sn isomers. On the basis of this multi-j $h_{11/2}\otimes d_{3/2}\otimes s_{1/2}$ configuration, one can also understand the origin of a higher seniority $v=5$, $({35/2}^+)$ isomer in odd-A Sn isotopes~\cite{jain20171}, which are only known for $^{123}$Sn~\cite{iskra2016}. The measurements for this $v=5,{35/2}^+$ isomer in neighboring $^{121,125,127}$Sn isotopes would help to test the seniority arguments for such a high-spin state. 

The same multi-j $h_{11/2} \otimes d_{3/2} \otimes s_{1/2} $ configuration could also describe the Q-moments of $v=1, {11/2}^-$ states in Sn isotopes~\cite{maheshwari20191}; however, the recent measurement numbers for the heavier Sn isotopes, particularly in $^{126,128,130}$Sn, are in contrast to the linear trend supporting a quadrupole trend instead~\cite{yordanov2020}. This may also be understood in terms of different rates of filling orbitals before and after $N=73$, the middle of the subshell involving $h_{11/2} \otimes d_{3/2} \otimes s_{1/2} $. The realistic non-degenerate generalized seniority scheme for many orbitals can improve the theoretical interpretation to explain this deviation of quadrupole moments from a linear trend.

\begin{figure}[!ht]
\includegraphics[width=0.8\textwidth]{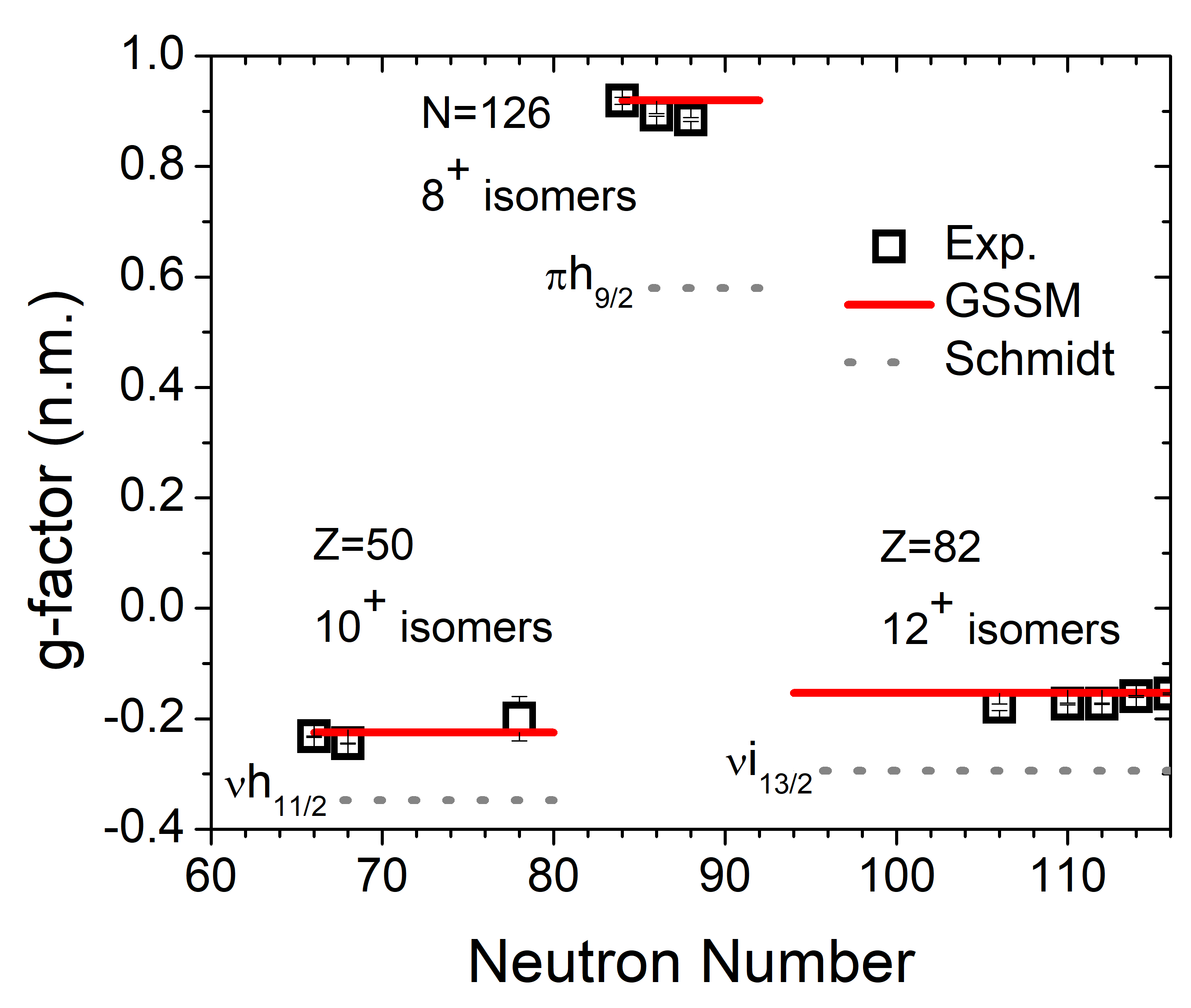}
\caption{\label{fig:gall}(Color online) {g$-$factor} 
variation of \textit{v} = 2 isomers in different semi-magic chains. GSSM refers to the estimates from the Generalized Seniority Schmidt Model. Exp. data have been taken from \cite{stone2014}.}
\end{figure}

The constant g-factor trend with an increasing neutron number for a given state is also a signature of good generalized seniority. It is imperative to interpret the ${11/2}^-$ states in Sn isotopes as the $v = 1$ pure-j seniority states with the $h_{11/2}$ unique-parity orbital in the $N = 50-82$ valence space. However, the experimental g-factors~\cite{stone2014} are substantially different from the Schmidt g-factor value for $h_{11/2}$ neutrons ($-0.348$ n.m.). Using the multi-j $h_{11/2} \otimes d_{3/2} \otimes s_{1/2} $ configuration in Schmidt formulation ($-0.225$ n.m.), referred to as GSSM, the g-factors for these ${11/2}^-$ states also come close to experimental data. This strongly suggests a mixed $ \tilde{j}= d_{3/2} \otimes s_{1/2}\otimes h_{11/2} $ configuration with shared occupancy for the $v=1,{11/2}^-$ states rather than a pure $h_{11/2}$~\cite{maheshwari2019}. The same GSSM estimated g-factor value could also explain the three experimental values known for the g-factors of $v=2, {10}^+$ isomers in $^{116,118,128}$Sn isotopes, as shown in Figure \ref{fig:gall}. It may be noted that there is no additional fitting involved in the estimation of GSSM values. The generalized seniority suggested configuration naturally takes care of the amount of spin quenching, which is usually required to explain the g-factor trends for these Sn isotopes. The nearly same g-factor values of the ${11/2}^-$ states and ${10}^+$ isomers suggests the similar degree of configuration mixing for both. The measured excitation energies for the ${10}^+$ (in even-A) and ${27/2}^-$ (in odd-A) isomers closely follow each other if one keeps the $0^+$ and ${11/2}^-$ states on equal footing, respectively. One can therefore strongly predict the g-factor values for the $v=3, {27/2}^-$ isomers in Sn isotopes to be of similar order for which no experimental data are known to date. As a result, the generalized seniority scheme consistently explains different spectroscopic features for these isomers in Sn isotopes.

The $6^+$ seniority isomers in neutron-rich Sn isotopes have been observed beyond the $N = 82$ shell closure \cite{simpson2014}. These seniority isomers serve as a useful probe for understanding the effective interaction in neutron-rich nuclei \cite{simpson2014,maheshwari2015}. The B(E2) measurements \cite{simpson2014} for the $6^+$ isomers in $^{134,136,138}$Sn isotopes do not follow the U-shape parabola, which is expected from single-j seniority arguments. This is because of the fact that the middle nucleus $^{136}$Sn with four neutrons active in $f_{7/2}$ can have two allowed $4^+$ states; one of them being $v=2$ while the other is $v=4$. This is similar to the case of $^{44}$Ca. The Berry phase at the middle of the $f_{7/2}$ orbital in $^{136}$Sn supports the allowed $\Delta v=4$ interaction between the ground $v=0, J=0$ and the $v=4, J=4$ states. This further results in the lowering of the $v=4, J=4$ state below the $6^+$ isomer, leading to an extra and faster $\Delta v=2$ decay branch for the $6^+$ isomer, resulting in the increased B(E2) value for the mid-shell nucleus, $^{136}$Sn. This has been achieved in shell model calculations by reducing the diagonal and non-diagonal neutron $f_{7/2}^2$ two-body matrix elements of realistic and effective interaction \cite{simpson2014,maheshwari2015}. It is also noted that the second $f_{7/2}$ (among nuclear single-particle levels) in the very neutron-rich Sn isotopes is surrounded by other higher-j orbitals and can also cause configuration mixing in the resulting wave functions for the seniority isomers \cite{jain20171}. This was not possible for first $f_{7/2}$ in Ca isotopes, because the other orbitals in \textit{fp}-valence space cannot mix with the wave functions of $6^+$ states.  

The spectroscopic approach to the extremely neutron-deficient doubly magic $^{100}$Sn has made substantial progress since the 1990s; see recent review~\cite{gorska2022}. The $6^+$ isomers are also known for a few particles above the $^{100}$Sn core. This is due to the activeness of neutron $g_{7/2}$ and $d_{5/2}$ orbitals altogether leading to the most-aligned $v=2$, $6^+$ state, as also shown in Figure \ref{fig:snevenenergy}. Low-lying excitations of $^{102-112}$Sn isotopes can be described to the unpaired neutrons in the $g_{7/2}$, $d_{5/2}$, and $h_{11/2}$. The possible $v=2$ configurations for the  yrast $6^+$ state are ${(g_{7/2} \otimes d_{5/2})}^2$, $g_{7/2}^2$, and $h_{11/2}^2$. The excitation energy of the last configuration is estimated to be twice the single-particle energy plus the pairing energy, which is too high to mix with the others. Using Schmidt values of $g_{7/2}$ and $d_{5/2}$, the calculated g-factors of the first two are $g_{g_{7/2} \otimes d_{5/2}}= - 0.07$ and $g_{g_{7/2}}= +0.43$ \cite{makishima1994}. The experimental known values in $^{108,110}$Sn isotopes come closer to the first value; however, the sign is not known in the case of $^{110}$Sn. The measured value in $^{106}$Sn is somewhat larger than the calculated estimate for the ${g_{7/2} \otimes d_{5/2}}$ configuration, but the sign supports this configuration. The $6^+$ isomer in $^{102}$Sn has recently been revisited for the lifetime and \textit{E}2 effective charge analysis~\cite{grawe2021}. Interesting possibilities of future experiments have been highlighted around $^{100}$Sn: to decipher the \textit{E}2 response of $^{100}$Sn and the similarity to $^{56}$Ni, to explore the proton decay out of super-deformed states in light Pd and Ru isotopes, and to investigate the proton and alpha decay from excited states of the lightest Sb and Te isotopes.  

\subsection{Pb Isotopes}

The isomerism is very well related to the occurrence of intruder and unique-parity high-j orbitals in a nuclear shell structure. The region around $^{208}$Pb is of special interest due to the proximity of many high-j orbitals~\cite{neyens2003}. The next intruder orbital after $h_{11/2}$ would be the $i_{13/2}$ orbital which becomes active in the region of neutron-deficient Pb isotopes with $N<126$. Their structure has been one of the most active subjects of nuclear physics research due to the possible coexistence of states with different shapes both experimentally and theoretically~\cite{heyde1983, bengtsson1989, aberg1990, wood1992, duppen2000, andreyev2000, julin2001, duguet2003, egido2004, guzman2004, frank2004, bender2004, grahn2006, wilson2010, heyde2011, nomura2012, delion2014, bree2014, julin2016, nomura2016, lalovic2018}. The ${13/2}^+$, ${33/2}^+$, and ${12}^+$ isomers dominated by the $i_{13/2}$ configuration are found regularly in odd-A and even-A $^{188-204}$Pb isotopes, respectively~\cite{maheshwari2021}. These $i_{13/2}$-dominated isomers can efficiently be used as a probe to differentiate the states of single-particle and collective nature, especially in the well-known region of shape coexistence for Pb isotopes. The ${13/2}^+$, ${12}^+$, and ${33/2}^+$ isomers can be deciphered as generalized seniority $v=1$, $v=2$ and $v=3$ isomers, respectively, using the multi-j $i_{13/2} \otimes f_{7/2} \otimes p_{3/2}$ configuration in Pb isotopes (for the mass range $A<200$); however, the dominance of the $i_{13/2}$ orbital is evident. Figure \ref{fig:BE2-GS}b exhibits the B(E2) variation for the $v=2, {12}^+$ isomers in Pb isotopes. The experimental information for these Pb isomers below $N=106$ is not available so far. It is very similar to the situation of $h_{11/2}$-dominated seniority isomers in Sn isotopes. 

The measured g-factors for these ${13/2}^+$, ${12}^+$, and ${33/2}^+$ isomers display a particle number-independent behavior as expected from generalized seniority but lie far away from the Schmidt value for a neutron $i_{13/2}$ orbital ($-0.295$ n.m.). As an example, we show the situation for the ${12}^+$ Pb isomers in Figure \ref{fig:gall}. The GSSM calculated g-factor values for the suggested multi-j $i_{13/2} \otimes f_{7/2} \otimes p_{3/2}$ configuration lie on the experimental data, which further supports the importance of configuration mixing in the origin of these $i_{13/2}$-dominated isomers \cite{maheshwari2019}. Since g-factor values for all the ${13/2}^+$, ${12}^+$, and ${33/2}^+$ isomers are nearly in the same order, one can expect a similar configuration mixing for the ${13/2}^+$ isomers (dominated by unique-parity intruder $i_{13/2}$ orbital) with generalized seniority $v=1$. \linebreak The ${13/2}^+$ states in Pb isotopes are predicted to emerge from the unique-parity $i_{13/2}$ orbital of $N=82-126$ valence space, which is just like the ${11/2}^-$ states in the Sn isotopes. Yet, it continues to follow the GSSM trend for the g-factor values, extending the applicability of the generalized seniority approach to $j=13/2$ for a particular set of states. Although the generalized seniority $v=1,{13/2}^+$, $v=2,{12}^+$, and $v=3,{33/2}^+$ isomers are nearly spherical, the quadrupole moments of these isomers are known in a few neutron-deficient Pb isotopes with $N>105$. Consequently, all the known Q-moment measurements \cite{stone2014} have positive signs and can be explained using the multi-j $i_{13/2} \otimes f_{7/2} \otimes p_{3/2}$ configuration \cite{maheshwari2021}. This can be understood in terms of the quasi-particle picture of the filling nucleons distributed among the active orbitals. The amount of configuration mixing is not dominating but is required to decipher the slower rate of filling nucleons in the $i_{13/2}$ orbital. This is evident due to the existence of ${13/2}^+$ isomers beyond the fourteen nucleons capacity of the $i_{13/2}$ orbital.  

\begin{figure}[!ht]
\includegraphics[width=12cm,height=10cm]{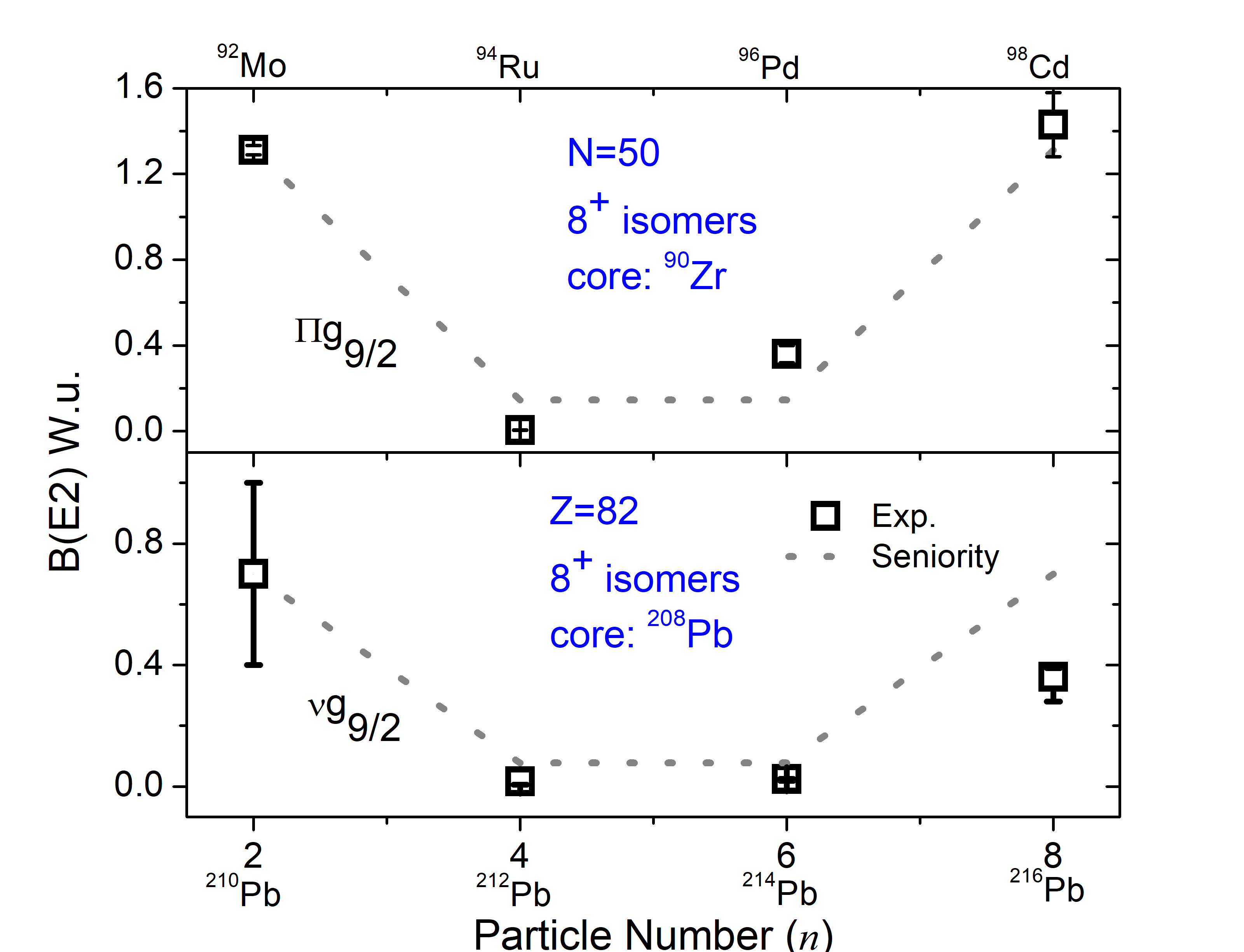}
\caption{\label{fig:PbN50}(Color {online}) 
A comparison of the experimental~\cite{ensdf} and seniority calculated B(E2) trends from the  g\textsubscript{9/2} orbital for the seniority \textit{v} = 2, 8\textsuperscript{+} isomers in both the \textit{N} = 50 isotones (upper panel) and Pb isotopes (lower panel).}
\end{figure}

The neutron-rich lead nuclei, with valence neutrons outside the doubly magic core $^{208}$Pb, serve as an ideal ground to study the seniority isomers due to the active second $g_{9/2}$ orbital and compare them with the first $g_{9/2}$-dominated isomers in lighter mass nuclei of $N=50$ isotones. The experimental access to this region of the nuclear chart has rather been difficult. Only the electromagnetic properties for the seniority isomers in $^{210}$Pb were known \cite{decman1983}. In 2012, Gottardo et al. \cite{gottardo2012} have reported the very first spectroscopic results on a number of neutron-rich Pb isotopes with $N > 126$ where the $8^+$ seniority isomers are systematically studied. The experimental~\cite{ensdf} and seniority calculated B(E2) trends for the $8^+$ isomers in even-A $^{210-216}$Pb isotopes are shown in Figure \ref{fig:PbN50}. These isomers can be understood in terms of a $v = 2$ configuration in the neutron $g_{9/2}$ orbital beyond the $^{208}$Pb core. With the exception of the $^{216}$Pb deviation, the seniority results largely explain the experimental B(E2) rates. The three-body forces has been used~\cite{gottardo2012} to explain the discrepancy between the seniority estimated and observed trends. For $^{210}$Pb, there is just one measured g-factor value ($-0.313$ n.m.), which is also considerably distant from the Schmidt value of neutron $g_{9/2}$ ($-0.426$ n.m.). This highlights the need for an extra mechanism to explain these $v=2,8^+$ seniority isomers in Pb isotopes. One of the ways could be to include the configuration mixing from the neighboring higher-j orbitals using the generalized seniority arguments. This may also be in line with the isotopic shift measurements for these neutron-rich Pb isotopes \cite{fricke2004}, requiring the wave functions to be spread out over various orbitals \cite{bhuyan2021}. The simple shell model calculations for $^{210}$Pb support the significant contributions in the $i_{11/2}$ orbital $(\sim 18 \%)$ besides the dominating $g_{9/2}$ orbital $(\sim 66 \%)$. 

The existence of ${21/2}^+$ isomers in odd-A Pb isotopes beyond $^{208}$Pb is also known. The known isomeric half-lives in $^{211,213}$Pb are 42 $ns$ and 0.26 $\mu s$, respectively, and are interpreted as seniority $v=3,g_{9/2}$-dominated isomers. The recent measurement in $^{213}$Pb~\cite{dobon2021} demonstrates that the $v=3$, ${21/2}^+$ isomer decays to two close-in-energy but dissimilar ${17/2}^+$ states. One of them has a seniority of $v = 3$, while the other has $v = 5$ and has no pair. The self-conjugate nature of $^{213}$Pb with five neutrons in $g_{9/2}$, where the particle--hole symmetry prohibits the mixing of two ${17/2}^+$ states owing to the Berry phase~\cite{dobon2021}, indicates the validity of seniority in the ${17/2}^+$ states. It would be interesting to search for similar isomers in a more heavier Pb isotope, $^{215}$Pb. The g-factor value for the ${9/2}^+$ ground state in $^{211,213}$Pb is also known and very similar to the g-factor of the $8^+$ isomer in $^{210}$Pb, suggesting their structure to be very similar except for the odd--even difference. The value is little far from the Schmidt neutron $g_{9/2}$ orbital, and it may require spin quenching due to extra mechanisms such as core excitation, configuration mixing, etc. Using the seniority arguments, one can strongly predict the g-factor of the ${21/2}^+$ isomers to be in similar order.  

\subsection{N = 28 Isotones}

The proton $f_{7/2}$ and corresponding $v=2$, $6^+$ isomers are also found in $N=28$ isotonic nuclei, $^{50}$Ti (0.4 $ns$), $^{52}$Cr (0.04 $ns$), $^{54}$Fe (1.22 $ns$)~\cite{ensdf}. The two-particle and two-hole symmetric nuclei, $^{50}$Ti (0.4 $ns$), and $^{54}$Fe (1.22 $ns$), display very similar features. The mass dependence in the $f_{7/2}^{(\pm 2)}$ matrix elements is visible in terms of isomeric locations. The $f_{7/2}^4$ nucleus, $^{52}$Cr has two low-lying $4^+$ states of different seniorities below the $6^+$ isomer. This changes the situation for the middle nucleus from the seniority expectations, as in the Ca isotopes. The g-factor for the ${6}^+$ isomer in $^{50}$Ti is known to be $+1.55$ n.m. which is quite near to the Schmidt value of pure proton $f_{7/2}$. The g-factor for the $6^+$ isomer in $^{54}$Fe is also known to be $1.37$ n.m. but without a sign. As per the seniority and single-j shell model arguments, it should be with a positive sign. Similar arguments can be given for the $6^+$ isomer in $^{52}$Cr.

\subsection{N = 50 Isotones}

It is generally known that the $8^+$ isomers in $N=50$ isotones $^{92}$Mo, $^{94}$Ru, $^{96}$Pd and $^{98}$Cd are seniority $v = 2$ isomers originating from the $g_{9/2}$ orbital. For these $8^+$ isomers in $N = 50$ isotones, Figure \ref{fig:PbN50} compares experimental \cite{atlas22} and seniority estimated B(E2) rates. The observed B(E2) trend may be quite accurately reproduced by the seniority results using $v = 2$ in the $g_{9/2}$ orbital, as shown in Figure \ref{fig:PbN50}. The observed g-factor for $^{92}$Mo, $^{94}$Ru, and $^{96}$Pd~\cite{stone2014}, which is nearly constant and extremely close to the Schmidt value for proton $g_{9/2}$, is also consistent with seniority expectations. This is due to the isolation of the proton g$_{9/2}$ orbital from its neighboring orbitals with no chances of significant mixing from neighboring orbitals in the resultant wave functions for these yrast $v=2,8^+$ isomers. One can, therefore, expect the similar order of g-factor for the $8^+$ isomer in $^{98}$Cd using seniority arguments. The influence of neighboring orbitals such as $p_{1/2},p_{3/2},f_{5/2}$ for these $g_{9/2}$-dominated isomers has been studied by Qi~\cite{qi2017}.

The recent measurements for these isotones~\cite{vidal2022,das2022} have also shown that seniority is largely conserved in the first proton $g_{9/2}$ orbital. This provides an interesting comparison of these $g_{9/2}$-dominated $8^+$ isomers in heavier Pb isotopes beyond $^{208}$Pb, as also shown in Figure~\ref{fig:PbN50}. In $N=50$ isotones, protons are the valence particles, while in Pb isotopes, neutrons are the valence particles. It may be noted that the $g_{9/2}$ orbital in the isotones is the first g-orbital; it is quite isolated and of unique parity from the neighboring orbitals. This is quite different to the situation in heavier Pb isotopes where $g_{9/2}$ is the second g-orbital and surrounded by neighboring higher-j orbitals.

Similar is the case for the $v=3, {21/2}^+$ isomers in odd-A $N=50$ isotones. The results of the proton $g_{9/2}$ seniority calculations closely match the observed trend~\cite{jain2021}. The measured value for $^{97}$Ag deviates from the seniority trend. No g-factor measurements are known for these $v=3$ isotonic isomers. Seniority suggests the g-factors of the $v=3$ isomers in odd-A isotones to be in the same order of the $v=2$ isomers in even-A.

\subsection{N = 82 Isotones}

The ${10}^+$ and ${27/2}^-$ isomers are known in the even-A and odd-A $N =82$ isotones $(Z \geq 66)$, respectively, and are interpreted as seniority $v=2$ and $v=3$ isomers arising from the $h_{11/2}$ protons~\cite{mcneill1989}. It is noted that the protons in $N=82$ isotones occupy the $g_{7/2}$, $d_{5/2}$, $h_{11/2}$, $d_{3/2}$, and $s_{1/2}$ orbitals in the $50$--$82$ nucleon space similar to the neutrons in Sn isotopes, allowing a crucial comparison between the two. The generalized seniority calculations are performed assuming the transition to be seniority preserving $\Delta v = 0 $ transitions and by fitting the $n = 2$ situation at $Z = 66$, $^{146}$Dy for the seniority $v = 2$ isomers. Such calculated results explain the resultant parabolic trend as shown in Figure~\ref{fig:BE2-GS}c. $Z = 64$ is considered as a core. The calculated B(E2) values using single-j $h_{11/2}$ with $\Omega=6$ also explain the experimental data very well up to $Z=70$, which is a different situation as compared to the ${10}^+$, $Z=50$ isomers, where the generalized seniority configuration of $h_{11/2} \otimes d_{3/2} \otimes s_{1/2}$ with pair degeneracy $\Omega=9$ gives the best fit. 
This may be due to the different proton single-particle energies for the active orbitals in $N=82$ isotones than to the neutron single-particle energies in Sn isotopes, although the involved orbitals are the same. No experimental information for these $N=82$ isomers exists  after $Z=72$. There are no moment measurements available for these ${10}^+$ and ${27/2}^-$ isomers in $N=82$ isotones which could provide additional structural information on the wave functions. In addition, \linebreak Astier et al. \cite{astier20121} has reported the high-spin structure of $N=82$ isotones having $Z=54-58$ where the ${10}^+$ isomers are considered to be arising from the $v=4$, $g_{7/2}$ and $d_{5/2}$ protons. 

The $8^+$ isomers are known to exist in neutron-rich $^{128}$Pd~\cite{watanbe2013} and $^{130}$Cd~\cite{jungclaus2007}. The more neutron-rich $^{126}$Ru and $^{124}$Mo are also expected to exhibit similar isomeric states. The $8^+$ isomer in the $^{128}$Pd, the r-process waiting-point nucleus, has a half-life of 5.8(8) $\mu s$ and is most likely associated to a maximally aligned $v=2$, $g_{9/2}$ configuration. While the first $2^+$ state is higher in $^{128}$Pd than $^{126}$Pd, the level sequence below the $8^+$ isomer in $^{128}$Pd is comparable to that in the $N = 82$ isotone $^{130}$Cd. The isomeric $E2$ transition in $^{128}$Pd is more hindered for the $8^+$ isomers than in $^{130}$Cd, which is consistent with the  seniority scheme. The existence of seniority isomers in these neutron-rich nuclei suggest the robust $N=82$ shell closure.

\subsection{N = 126 Isotones}

The ${9/2}^-$, $8^+$ and ${21/2}^-$ isomers in $N=126$ isotones (beyond $^{208}$Pb) are generally described using $h_{9/2}$ due to their seniority such as B(E2) rates \cite{ressler2005}. In contrast to this, the measurements on $^{216}$Th isotopes \cite{zhang2019} lead to a distinct behavior and reveal a very low B(E2) value. The experimental g-factor values of these ${9/2}^-$, $8^+$ and ${21/2}^-$ isomers \cite{stone2014} likewise deviate significantly from the proton $h_{9/2}$ Schmidt value. Finding a consistent configuration to explain both isomeric decays and moments has been addressed using generalized seniority~\cite{maheshwari2022}. The role of configuration mixing for all the three isomers is found to be crucial. These $h_{9/2}$-dominated ${9/2}^-$, $8^+$ and ${21/2}^-$ isomers are deciphered as $v=1$, $v=2$ and $v=3$ isomers, respectively, from the multi-j proton $h_{9/2} \otimes f_{7/2} \otimes i_{13/2}$ configuration. The best calculated results are found with configuration mixing corresponding to $\Omega=7$ which lie on the experimental data. The B(E2) results for the $8^+$ isotonic isomers are shown in Figure \ref{fig:BE2-GS}d. The g-factor results using GSSM are compared with the experimental data and Schmidt value for these isomers in Figure \ref{fig:gall}. This choice of $\Omega=7$ is based on the limited mixing of $f_{7/2} \otimes i_{13/2}$ orbitals to the resulting total $h_{9/2}$ dominated wave functions. This is analogous to the non-degenerate description of various active orbitals. The shell model calculations for these isomers also validate the generalized seniority interpretation for $N=126$ isotonic isomers~\cite{maheshwari2022}. Such phenomenological calculations become important when the other microscopic model calculations become cumbersome and can help in predictions. 

The activeness of proton $h_{9/2}$ and neutron $g_{9/2}$ beyond $^{208}$Pb provides an interesting possibility to have isomers with same $v$ and $J$, $v=2, 8^+$ isomers in even-A semi-magic Pb isotopes or $N=126$ isotones. Their nuclear surroundings, especially in terms of neighboring orbitals, is quite different to each other, and so are their g-factors. The comparison between $^{216}$Pb and $^{216}$Th would be able to distinguish the two-body interaction and role of involved nucleonic configurations.   

\subsection{Cd and Te Isotopes}

No nucleus is deformed unless there are at least four valence nucleons, and a nucleus must be deformed if there are at least 10 valence nucleons of each type \cite{casten}. This indicates the goodness of generalized seniority up to two-particles/holes away from the semi-magicity. For this, one can try to focus on the trends of Q-moments, which provides a measure of the shape deviation from the sphere. The simple linear behavior of Q-moments for the ${11/2}^-$ states in the Cd isotopes beyond the $h_{11/2}$ orbital, many of which are isomers, has been noted by Yordanov et al. \cite{yordanov2018}. The generalized seniority scheme with a mixed configuration for the ${11/2}^-$ states in Cd and Te isotopes~\cite{maheshwari2019} could explain the linear incremental trend of experimental Q-moments~\cite{stone2014} even beyond the $h_{11/2}$ orbital. For all three of the isotope chains, $^{123}$Sn ($N=73$), $^{121}$Cd ($N=73$) and $^{125}$Te ($N=73$), the Q-moment shifts from a negative to a positive value in the middle $N = 73$. The theoretical values are derived by fitting one of the experimental data to the $h_{11/2} \otimes d_{3/2} \otimes s_{1/2}$ configuration (which fixes the involved reduced matrix elements for the $E2$ operator). Beyond $N=64$, the $g_{7/2}$ and $d_{5/2}$ orbitals may be assumed to be fully filled for the neutron-rich Cd, Sn, and Te isotopes, since they become active as soon as neutrons start to fill the $N=50$--$82$ and are lower in energy than the $h_{11/2}, d_{3/2}$, and $s_{1/2}$ orbitals. Lighter Cd isotopes (before $N=64$), in particular $^{109,111}$Cd, have Q-moment values that are closer to the generalized seniority results with a $v =1$ and $\Omega=12$ configuration. Such computed Q-moment values start to vary from the measured behavior at $^{113}$Cd with $N=65$ and eventually deviate for heavier Cd isotopes.

This further supports the $N = 64$ sub-shell closure; nevertheless, for the lighter $(N<64)$ Cd isotopes, the potential $d_{5/2}$ orbital mixing cannot be completely ruled out \cite{maheshwari20191}. All three Cd, Sn, and Te isotopic chains have a comparable range of Q-values for the ${11/2}^-$ states between \textit{N} = 65 and 81, confirming a similar structure of these states as one moves from Cd (two proton-holes), Sn (the proton closed-shell), and Te (two proton-particles) isotopes. Therefore, these analyses provide a generalized seniority explanation for the structural evolution within and around the $Z = 50$ closed shell. It should be highlighted that the computed Q-moment trend was unable to account for the data for the isotopes $^{131,133}$Te $(N = 79, 81)$ and $^{129,131}$Sn $(N = 79, 81)$. It is necessary to conduct additional research using the generalized seniority scheme with both protons and neutrons, and non-degenerate multi-j orbitals, as recent measurements of Q-moments \cite{yordanov2020} for the ${11/2}^-$ states in $^{129,131}$Sn isotopes show a quadratic trend rather than a linear trend (seen in Cd isotopes). Since merely $h_{11/2}$ cannot account for the entire observed trend, the multi-j description using a generalized seniority scheme is discovered to be crucial. Future measurements could validate this, especially for higher Te isotopes. Additionally, it is found that the g-factor trends for the ${11/2}^-$ states in all three isotope chains are almost particle number independent, lying close to one another \cite{maheshwari20191}.

The high-spin ${10}^+$ isomers are also known to exist in $^{126,128,130,132}$Te isotopes. These isomers are also dominated by the neutron $h_{11/2}^2$ configuration, which is very similar to the case of Sn isotopes in this mass region. Their B(E2)s do not follow the U-shape of seniority due to competing proton components in their resulting wave functions. Although the two available protons above the $Z=50$ closed shell in Te isotopes could only mix the $v=2$ wave functions up to $6^+$ states, the competing proton components in the wave functions of ${10}^+$ states from shell model calculations \cite{astier2014,kumar2015} strongly suggest the mixing of $v=4$ components ($v=2$ protons coupled to $v=2$ neutrons). If this is the case, seniority is not able to dominate the trend of B(E2) rates, which significantly depends upon the \textit{E}2 coupling of nuclear wave functions for the ${10}^+$ and $8^+$ states. Although the possibility of ${10}^+$ isomers in more lighter Te isotopes, $^{122,120,118}$Te isotopes cannot be ruled out and would be worthwhile to visit. On the basis of two-proton particles/holes symmetry, one can expect to have similar ${10}^+$ isomers in Cd isotopes with neutrons in $50$--$82$ valence space. In $^{128}$Cd $(N=80)$, the ${10}^+$ isomer exists. However, what will happen in more lighter $^{126,124,122..}$Cd isotopes would be able to infer the evolution of single-particle orbitals as well as the two-body interaction. A comparison of neutron--neutron interaction in $^{128}$Cd, $^{132}$Te isotopes to the $^{130}$Sn would enable deciphering the same. Similar results may be expected for the higher seniority $v=4, {15}^-$ isomers in Cd and Te isotopes. Such ${15}^-$ isomers are already known in $^{128}$Cd and $^{130}$Te, similar to the Sn isotopes~\cite{atlas22}.  

\subsection{Hg and Po Isotopes}

The ${13/2}^+$, ${12}^+$, and ${33/2}^+$ isomers are also known to occur in Hg and Po isotopes similar to the Pb isotopes. The generalized seniority calculations could explain their various spectroscopic properties including decay probabilities, g-factors and Q-moments (wherever available) using a consistent multi-j configuration~\cite{maheshwari2021}. These ${13/2}^+$, ${12}^+$, and ${33/2}^+$ isomers in Hg and Po isotopes are also assigned the respective generalized seniority of $v=1$, $v=2$, and $v=3$, which is very similar to the semi-magic chain of Pb isotopes. Despite the two-proton holes/particles involved in Hg and Po isotopes and related complexities, these $i_{13/2}$-dominated isomers support regular trends of magnetic dipole (g-factor) and electric quadrupole (Q) moments. The multi-j $i_{13/2} \otimes f_{7/2} \otimes p_{3/2}$ configuration is able to explain the transition probabilities and moments consistently because of many-body symmetries in terms of generalized seniority. The GSSM estimate of g-factor values using a multi-j configuration explains the measured data in and around Pb isotopes. Predictions have been made requiring dedicated future experimental studies to understand the systematics of such seniority isomers in a better way for this neutron-deficient region.

In addition, the $v=2, 8^+$ isomers due to the $g_{9/2}$ orbital is known for the neutron-rich $^{208,210}$Hg and $^{212}$Po isotopes \cite{dahan2009}, similar to the case of neutron-rich $^{210,212,214,216}$Pb isotopes \cite{decman1983,gottardo2012}. Similar isomers may be expected for the $^{212,214}$Hg isotopes, where high-spin spectroscopy is not yet known. In Hg isotopes, two proton holes are in the $d_{3/2}$ and $s_{1/2}$ orbitals, which would behave as the proton spectators for the structure of yrast $v=2, 8^+$ states. Hence, it is quite probable to find the $8^+$ isomers in these neutron-rich Hg isotopes. The search for these neutron-rich seniority isomers in $^{214,216,218}$Po isotopes would be equally interesting where two protons in $h_{9/2}$ and two neutrons in $g_{9/2}$ start to compete for the structure of yrast $v=2,8^+$ states. The moment measurements along with lifetimes would be able to conclude the situation.   

\subsection{N = 48 and N = 52 Isotones}

The region near $^{100}$Sn is crucial for the nuclear structure around $N \sim Z$ and astrophysical processes~\cite{grawe2006,fastermann2013,park2017,gorska2022}. The two holes in a neutron $g_{9/2}$ orbital for $N=48$ isotones lead to the possibility of $8^+$ isomers. However, the valence protons would also start to occupy the $g_{9/2}$ orbital near $^{88}$Zr, leading to an additional and competing proton $8^+$ state in $N=48$ isotonic nuclei with $Z=40$--$50$. It may be noted that the two or four particles/holes in the $g_{9/2}$ orbital leads to only one $8^+$ state. So, the occurrence of two $8^+$ states would most likely happen when both the proton $g_{9/2}^2$ and neutron $g_{9/2}^2$ configurations become active. 

\begin{figure}[!ht]
\includegraphics[width=0.8\textwidth]{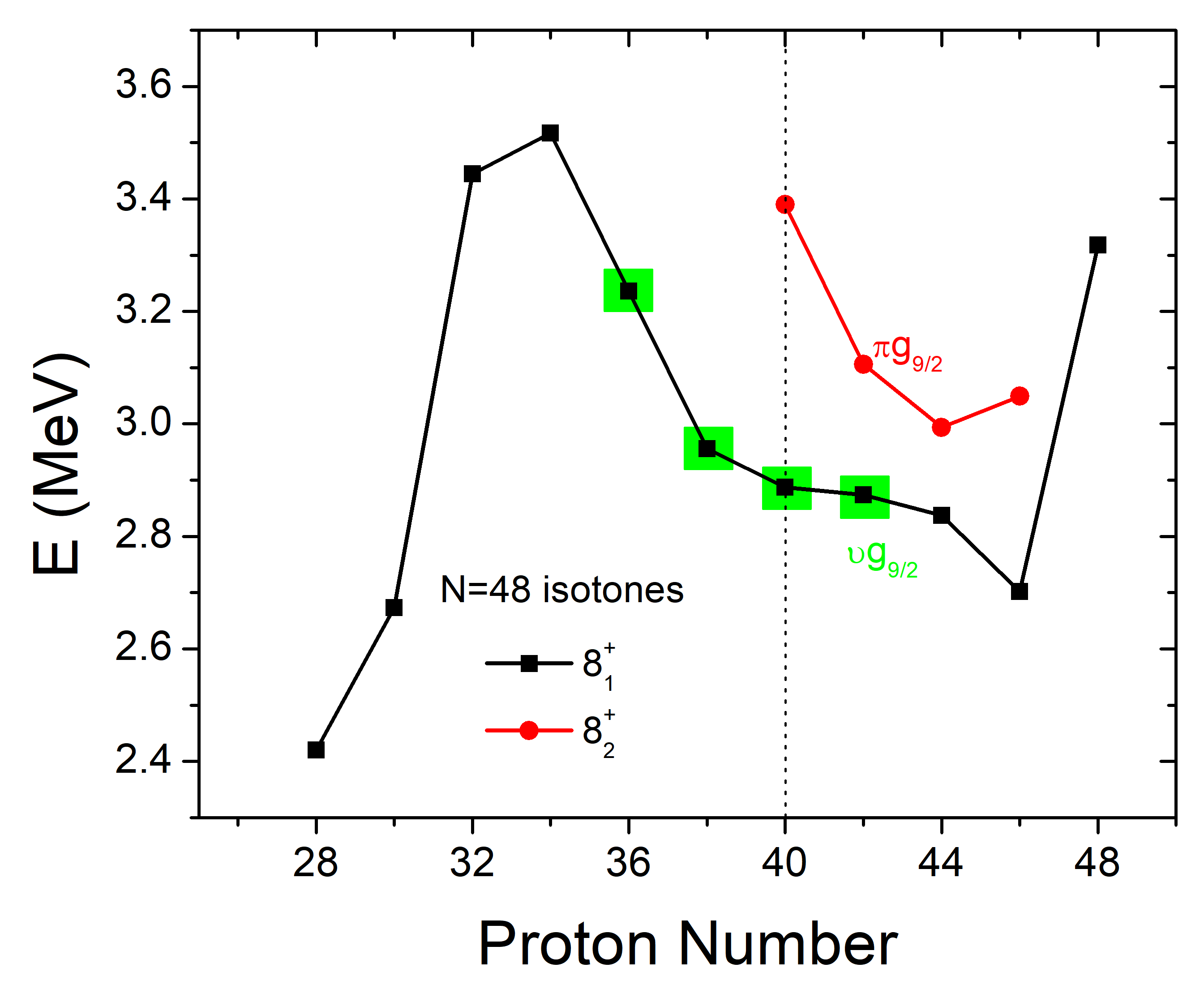}
\caption{\label{fig:8isomer}(Color {online}) Empirical energy systematics for the yrast and yrare 8\textsuperscript{+} states in \textit{N} = 48 isotones. The isotonic nuclei, where magnetic moment measurements for these $8_1^+$ isomers are known, are shown with green shade. The negative sign for these moments strongly supports the neutron $g_{9/2}^{-2}$ configuration.}
\end{figure}

The $8^+$ isomers are known to exist in $N=48$ isotones for even $Z$ ranging from $Z=28-44$. Figure \ref{fig:8isomer} represents the systematics of measured energies for the two $8_1^+, 8^+_2$ states in $N=48$ isotones. As expected, there are two $8^+$ states known for isotonic nuclei, $^{88}$Zr, $^{90}$Mo, $^{92}$Ru, and $^{94}$Pd~\cite{atlas22}. No half-lives are known for the two $8^+$ states in $^{94}$Pd, although the situation looks quite analogous to the $^{92}$Ru~\cite{atlas22,ensdf}. For the $N=48$ isotonic nuclei with $Z<40$, the active protons occupy lower lying $fp$ orbitals leading to mixed wave functions for the $0^+$ to $4^+$ states but not for the $6^+,8^+$ states. The g-factors for the $8_1^+$ isomers are also known for even-$Z$ ranging from $Z=36$--$42$, $^{84}$Kr $(-0.246$ n.m.), $^{86}$Sr $(-0.241$ n.m.), $^{88}$Zr $(-0.226$ n.m.) and $^{90}$Mo $(-0.174$ n.m.), as shown in Figure \ref{fig:8isomer} using green shades. All the moments are negative in sign, clearly indicating these $8_1^+$ isomers to be arising from two neutron holes in $g_{9/2}$. However, the experimental value remains quite far from the Schmidt value of $-0.426$ n.m. for the neutron $g_{9/2}$ and requires a spin attenuation factor to be in the order of 0.5 to explain the measurement.   
In $N=52$ isotones, the $8^+$ isomers are known in $^{94}$Mo and $^{100}$Cd with respective half-lives of 98 and 62 nanoseconds~\cite{atlas22}. In both cases, the $8^+$ isomers decay by two \textit{E}2 gammas to the two lower-lying $6^+$ states. The magnetic moment for the $8^+$ isomer in $^{94}$Mo is known as $+10.46(7)$ n.m. \cite{stone2014}: the positive sign clearly supporting the proton $g_{9/2}^2$ configuration. The magnetic moment value for the $8^+$ isomer in $^{100}$Cd is known to be $9.9(5)$ n.m. without the sign~\cite{stone2014}. The analogous situation for both $^{94}$Mo and $^{100}$Cd $N=52$ isotonic nuclei strongly suggests the same configuration for these $8^+$ isomers, and so is the positive sign for the magnetic moments. The measured magnetic moments are near to Schmidt values of proton $g_{9/2}$ with a spin quenching in the order of 0.8. The $8^+$ state in $^{96}$Ru and $^{98}$Pd also exists with respective half-lives of 11 and 66 picoseconds \cite{ensdf}. Both of them decay by \textit{E}2 transition to the lower-lying $6^+$ state. In the first approximation, the low-lying states can be explained by the coupling of two neutrons in $d_{5/2}$ or $g_{7/2}$. However, the additional coupling of valence protons and their interaction with neutrons becomes crucial in explaining the real situation. 

The region near $^{100}$Sn is crucial for the nuclear structure around $N \sim Z$ and astrophysical processes~\cite{grawe2006,fastermann2013,park2017,gorska2022}. The two holes in the neutron $g_{9/2}$ orbital for $N=48$ isotones lead to the possibility of $8^+$ isomers. However, the valence protons would also start to occupy the $g_{9/2}$ orbital near $^{88}$Zr and lead to an additional and competing proton $8^+$ state in $N=48$ isotonic nuclei with $Z=40-50$. It may be noted that the two or four particles/holes in the $g_{9/2}$ orbital leads to only one $8^+$ state. So, the occurrence of two $8^+$ states would most likely happen when both proton $g_{9/2}^2$ and neutron $g_{9/2}^2$ configurations become active. 

The $8^+$ isomers are known to exist in $N=48$ isotones for even $Z$ values in the range of $Z=28$--$44$. Figure \ref{fig:8isomer} represents the systematics of measured energies for the two $8_1^+, 8^+_2$ states in $N=48$ isotones. As expected, there are two $8^+$ states known for isotonic nuclei, $^{88}$Zr, $^{90}$Mo, $^{92}$Ru, and $^{94}$Pd~\cite{atlas22}. No half-lives are known for the two $8^+$ states in $^{94}$Pd, although the situation looks quite analogous to the $^{92}$Ru~\cite{atlas22,ensdf}. For the $N=48$ isotonic nuclei with $Z<40$, the active protons occupy lower lying $fp$ orbitals leading to mixed wave functions for the $0^+$ to $4^+$ states but not for the $6^+,8^+$ states. The g-factors for the $8_1^+$ isomers are also known for even-$Z$ ranging from $Z=36-42$, $^{84}$Kr $(-0.246$ n.m.), $^{86}$Sr $(-0.241$ n.m.), $^{88}$Zr $(-0.226$ n.m.) and $^{90}$Mo $(-0.174$ n.m.), as shown in Figure \ref{fig:8isomer} using green shades. \linebreak All the moments are negative in sign, clearly indicating these $8_1^+$ isomers to be arising from two neutron holes in $g_{9/2}$. However, the experimental value remains quite far from the Schmidt value of $-0.426$ n.m. for the neutron $g_{9/2}$ and requires the spin attenuation factor to be in the order of 0.5 to explain the measurement.   
In $N=52$ isotones, the $8^+$ isomers are known in $^{94}$Mo and $^{100}$Cd with respective half-lives of 98 and 62 nanoseconds~\cite{atlas22}. In both cases, the $8^+$ isomers decay by two \textit{E}2 gammas to the two lower lying $6^+$ states. \linebreak The magnetic moment for the $8^+$ isomer in $^{94}$Mo is known as +10.46(7) n.m. \cite{stone2014}; the positive sign clearly supports the proton $g_{9/2}^2$ configuration. The magnetic moment value for the $8^+$ isomer in $^{100}$Cd is known to be 9.9(5) n.m. without the sign~\cite{stone2014}. The analogous situation for both $^{94}$Mo and $^{100}$Cd $N=52$ isotonic nuclei strongly suggests the same configuration for these $8^+$ isomers and so is the positive sign for the magnetic moments. The measured magnetic moments are near to Schmidt values of proton $g_{9/2}$ with a spin quenching in the order of 0.8. The $8^+$ state in $^{96}$Ru and $^{98}$Pd also exist with respective half-lives of 11 and 66 picoseconds \cite{ensdf}. Both of them decay by \textit{E}2 transition to the lower lying $6^+$ state. In the first approximation, the low-lying states can be explained by the coupling of two neutrons in $d_{5/2}$ or $g_{7/2}$. However, the additional coupling of valence protons and their interaction with neutrons becomes crucial in explaining the real situation.   

\subsection{N = 80 and N = 84 Isotones}

In $N=80$ isotones, the ${10}^+$ isomers exist in two nuclei beyond $Z>64$, i.e., $^{146}$Dy and $^{148}$Er. The ${10}^+$ isomer in $^{146}$Dy decays by two \textit{E}3 gammas of 127 and 416 keV \cite{atlas22}. The ${(10)}^+$ isomer in $^{148}$Er decays by E2 transition to the lower-lying $(8^+)$ state. In both isotonic nuclei, the ${(10)}^+$ isomer is known to be arising as a $v=2$ isomer with two neutron holes in $h_{11/2}$~\cite{gui1982,nolte1982}. For $^{126}$Pd $(N=80)$, two new isomers are known with half-lives of 0.33(4) and 0.44(3) $\mu s$~\cite{atlas22}. Similar isomers are known in $^{128}$Cd with half-lives of 270 and 12 nanoseconds~\cite{atlas22}. These two isomers have been assigned tentative spins and parities as $(5^-)$ and $(7^-)$, respectively. The ${10}^+, v=2$ isomer due to the most-aligned two-neutron holes in $h_{11/2}$ is also known in $^{126}$Pd and $^{128}$Cd, $N=80$ isotonic nuclei~\cite{atlas22} and references therein. One can, therefore, expect the existence of these isomers in a more neutron-rich $N=80$ isotonic nucleus, $^{124}$Ru. No high-spin spectroscopy is available for this limiting and extremely neutron-rich nucleus.      

In $N=84$ isotones (with $Z>64$), the ${10}^+$ states also exist almost at the similar energies but without a half-life (i.e., no extra hindrance reported so far) except for the case of $^{150}$Dy having a ${10}^+$ isomer with a 1.1 \textit{ns} half-life~\cite{ensdf}. It may be noted that the two-neutron particles above $N=82$ also start to compete with the lower-lying configurations for $N=84$ isotones in addition to the valence protons from $64$--$82$ space. Consequently, the $8^+$ isomers exist in $^{152}$Er, $^{154}$Yb, $^{156}$Hf, and $^{158}$W isotonic nuclei arising from the two neutron particles above $N=82$ shell closure dominated by an $h_{9/2} \otimes f_{7/2}$ configuration~\cite{atlas22,ensdf}.

\subsection{N = 124 and N = 128 Isotones}

The $h_{9/2}$ dominated $v=2, 8^+$ isomers in $N=124$ isotones above the $^{208}$Pb core follow a U-shape B(E2) parabola as in the case of $N=126$ isotones and support the mixing of neighboring higher-j orbitals in a limited amount~\cite{maheshwari2022}. In $N=128$ isotones, the $8^+$ isomers also exist from $Z=84-90$. Among which, the isomers at $Z=84,86$ are mainly dominated by the two-protons configuration while the isomers at $Z=88,90$ are mainly dominated by two-neutrons configuration going away from the U-shape parabola. This is due to the availability of two-neutron particles above the neutron closed shell in these isotones along with the active proton valence space and can be confirmed by the g-factor value of $^{216}$Ra $(Z=88, N=128)$ \cite{stone2014}. Interestingly, the first active neutron orbital above $N=126$ is again a $j=9/2$ orbital, that is, $g_{9/2}$. This situation is similar to the case of Pb isomers beyond $^{208}$Pb, where the valence neutrons start to occupy the same $g_{9/2}$ above $N=126$. 

\section{Predictions and Open Issues}\label{sec5}

The effectiveness of seniority and generalized seniority in explaining the isomerism, particularly in heavier mass nuclei, is highly encouraging. Future experiments to evaluate the goodness of seniority may refer to uncharted regions of nuclei. Below are some predictions made from the perspective of generalized seniority:

\begin{enumerate}
    \item The measurement gaps in the systematics of seniority isomers require special attention from the experimentalists, since this missing piece of information would be crucial for the theoretical developments. The presence of $v=2, 8^+$ isomers in neutron-rich $^{128}$Pd and $^{130}$Cd, for example, strongly suggests the same $v=2$ isomers in more neutron-rich $^{126}$Ru and $^{124}$Mo, $N=82$ isotonic nuclei. Similarly, future experimental data on the $v=2, {10}^+$ isomer in neutron-rich $^{124}$Ru $(Z=44,N=80)$ (due to two neutron holes in $h_{11/2}$) will be immensely helpful in understanding the function of pairing in such a limiting and exceedingly neutron-rich nucleus. The existence of ${10}^+,v=2$ isomers in $^{128}$Cd also suggests the dominance of the neutron $h_{11/2}^{-2}$ configuration for the lower lying $8^+$ state. Any information on the $8^+$ state due to two proton $g_{9/2}$ holes in $^{128}$Cd $(Z=48, N=80)$ would be equally important for assessing the competitiveness between proton and neutron two-body configurations. Similar investigation for the $8^+$ states in $^{126}$Pd would also be encouraging to understand the structural evolution.
    \item The particle-number independent variation of the magnetic moments for the good (generalized) seniority states can be used to predict the g-factors for the gaps in measurements. For example, the g-factor for the $6^+$ isomer in $^{46}$Ca should be in the similar order to the g-factor of the $6^+$ isomer in $^{42}$Ca. Despite the fact that the two $6^+$ states in $^{44}$Ca have different seniorities, the g-factor should be equal to neutron $f_{7/2}$ owing to the pure-j configuration. Similarly, the g-factor in $^{70,76}$Ni isotopes for the $8^+$ isomers should be comparable and, if measured, would represent the nature of the implicated neutron $g_{9/2}$ orbital. The same can be said for the $8^+$ states in $^{72,74}$Ni isotopes, which are not isomeric due to the additional and permitted decay branch. Similar would be true for the seniority isomers in medium to heavy mass nuclei such as the ${10}^+$ isomers in $^{120,122,124,126,130}$Sn isotopes, the ${12}^+$ isomers in $^{190}$Pb, and lighter $^{186,184,182,..}$Pb isotopes. The same is true for the g-factors of seniority isomers in various isotonic chains.
    \item If the isomers have the same origin and only differ in terms of an extra odd-particle, the g-factor of odd-A isomers will be in the same order as that of even-A isomers for a given isotopic or isotonic chain. The g-factor for the ${27/2}^-$ isomers in odd-A $^{117,119,121,123,125,127,129}$Sn isotopes, for example, would be of the same order as the ${10}^+$ isomers in even-A Sn isotopes. The same will hold true for the odd-A ${33/2}^+$ isomers in lighter odd-A $^{183,185,187,189,191}$Pb isotopes due to their similarity to the neighboring even-A ${12}^+$ isomers. Similarly, the g-factor for the ${10}^+$ and ${27/2}^-$ isomers in respective even-A $Z=66,68,70,72$, and odd-A $Z=67,69,71,73$, $N=82$ isotones should be almost equal to each other. The Schmidt value for proton $h_{11/2}$ is $+1.42$ n.m., although the GSSM estimate for the mixed proton $h_{11/2}\otimes d_{3/2}\otimes s_{1/2}$ configuration is +1.27 n.m. Such future moment measurements would provide the complete understanding of a nuclear structure for these isotonic isomers.  
    \item To address the similarities and differences in the behavior of ${11/2}^-$ states in Cd, Sn and Te isotopes, the Q-moment measurements in heavier Te isotopes are of current experimental interest, particularly when similar measurements for Cd and Sn isotopes are now known with great precision at the ISOLDE facility~\cite{yordanov2018,yordanov2020}. Since the $v=1,{11/2}^-$ states are found to occur quite regularly in Cd, Sn and Te isotopes for the range of $N=65-81$, one can expect the higher seniority isomers such as $v=2,{10}^+$, $v=3,{27/2}^-$, $v=4,{15}^-$ in the Cd and Te isotopes, similar to the Sn isotopes.
    \item The lack of experimental data on \textit{E}2 decay properties of the first $2^+,4^+$ states below the seniority isomers in $N=50$ isotones prevents a conclusion on the seniority conservation in $j=9/2$ from being established. Similarly, in other heavier mass regions, firm \textit{E}2 assignments below the most-aligned seniority isomer are not yet available. The \textit{E}2 properties for the states below the neutron-rich $6^+$ seniority isomers in Sn isotopes beyond $^{132}$Sn and the states below the $8^+$ seniority isomers in Pb isotopes beyond $^{208}$Pb, for example, will undoubtedly contribute to realistic and effective nuclear shell model interactions. To fully comprehend the neutron--neutron/proton--proton as well as neutron--proton two-body matrix elements, comprehensive spectroscopic information for isomers and states below isomers in two-particles/holes nuclei with respect to semi-magic nuclei such as Cd and Te isotopes, Hg and Po isotopes, $N = 48$, 52 isotones, $N = 80$, 84 isotones, $N = 124$, 128 isotones is necessary.   
    \end{enumerate}

\section{Conclusions}\label{sec6}

We have provided an overview of seniority and generalized seniority isomers in different nuclear mass regions. Such isomers act remarkably similar to one another in completely distinct valence spaces as a result of the symmetries at play. Although the nuclear environment can cause configuration mixing in the resulting wave functions of these isomers, it has been discovered that the near goodness of generalized seniority can govern the evolution of their spectroscopic features. These seniority isomers are especially helpful in determining the rigidity of spherical magic numbers, comprehending the development of single-particle energies, and two-body interactions, especially the short-range pairing interaction. Simple seniority and generalized seniority explanations in different mass regions become quite essential to map the boundaries between the single-particle and collective motion of nucleons. These investigations are especially crucial for the nuclei where the microscopic calculations are quite involved. 

The fact that the deformation does not only arise from more number of available valence nucleons outside the closed shells but must involve both valence protons and neutrons specifically is also the consequence of the seniority picture. As soon as we move away by four or more nucleons from the semi-magic nuclei, the collective excitations begin to occur along with the good seniority states. Understanding the pace at which collective excitations start to dominate the single-particle picture would be fascinating. This will affect the isomeric transition's hindrance mechanism as well. K-isomers are one of these collective isomers in axially deformed nuclei; they are also associated with broken pair excitations in deformed nuclei, which are equivalent to seniority in spherical nuclei. It may also be possible to predict a relationship between distinct groups of isomers.

\vspace{6pt} 




\section*{Acknowledgments}The financial support 
from the Croatian Science Foundation and the \'Ecole Polytechnique F\'ed\'erale de Lausanne, under the project TTP-2018-07-3554 ``Exotic Nuclear Structure and Dynamics", with funds of the Croatian-Swiss Research Programme, is gratefully acknowledged.




\end{document}